\documentclass[%
reprint,
showpacs,preprintnumbers,
 amsmath,amssymb,
 aps,
 pre,
floatfix,
]{revtex4-1}

\usepackage{graphicx}
\usepackage{dcolumn}
\usepackage{bm}
\usepackage{color}
\usepackage{hyperref}

\providecommand{\vektor}[1]{\ensuremath{\boldsymbol{#1}}}
\providecommand{\Tensor}[1]{\ensuremath{\boldsymbol{#1}}}

\newcommand{\Ttot}{\ensuremath{\mathrm{T}}}
\newcommand{\Jel}{\ensuremath{\mathrm{J}}}
\newcommand{\Deform}{\ensuremath{\boldsymbol{\mathrm{D}}}}

\newcommand{\Imat}{\ensuremath{\boldsymbol{\mathrm{I}}}}
\newcommand{\Mmat}{\ensuremath{\boldsymbol{\mathrm{M}}}}
\newcommand{\Cmat}{\ensuremath{\boldsymbol{\mathrm{A}}}}
\newcommand{\Jacobian}{\ensuremath{\boldsymbol{\mathrm{J}}}}
\newcommand{\velocity}{\ensuremath{\boldsymbol{\mathrm{v}}}}
\newcommand{\unitvector}{\ensuremath{\hat{\boldsymbol{\mathrm{x}}}}}
\newcommand{\Ttensor}{\ensuremath{\boldsymbol{\mathrm{T}}}}

\newcommand{\VStensor}{\ensuremath{\boldsymbol{\sigma}}}

\begin{document}


\title{Time-delayed feedback control of shear-driven micellar systems}

\author{Benjamin von Lospichl}%
 \email{blospichl@itp.tu-berlin.de}%
\altaffiliation[present address: ]{Stranski-Laboratorium f\"{u}r Physikalische und Theoretische Chemie, Institut f\"{u}r Chemie, Sekretariat TC 7, Technische Universit\"{a}t Berlin, Stra\ss e des 17. Juni 124, D-10623 Berlin, Germany}%
\affiliation{Institut f\"{u}r Theoretische Physik, Sekretariat EW 7-1, Technische Universit\"{a}t Berlin, Hardenbergstra\ss e 36, D-10623 Berlin, Germany
}%
\author{Sabine H. L. Klapp}%
 \email{klapp@physik.tu-berlin.de}
\affiliation{Institut f\"{u}r Theoretische Physik, Sekretariat EW 7-1, Technische Universit\"{a}t Berlin, Hardenbergstra\ss e 36, D-10623 Berlin, Germany
}%

\date{\today}

\begin{abstract}
Suspensions of elongated micelles under shear display complex non-linear behaviour including shear banding, spatio-temporal oscillatory patterns and chaotic response. Based on a suitable rheological model [S. M. Fielding and P. D. Olmsted, {\it Phys.~Rev.~Lett.} {\bf 92}, 084502 (2004)], we here explore possibilities to manipulate the dynamical behaviour via closed-loop (feedback) control involving a time delay $\tau$. The model considered relates the viscoelastic stress of the system to a structural variable, that is, the length of the micelles, yielding two time- and space-dependent dynamical variables $\xi_1$, $\xi_2$. As a starting point we perform a systematic linear stability analysis of the uncontrolled system for (i) an externally imposed average shear rate and (ii) an imposed total stress, and compare the results to those from extensive numerical simulations. We then apply the so-called Pyragas feedback scheme where the equations of motion are supplemented by a control term of the form $K\left(a(t)-a(t-\tau)\right)$ with $a$ being a measurable quantity depending on the rheological protocol. For the choice of an imposed shear rate, the Pyragas scheme for the total stress reduces to a non-diagonal scheme concentrating on the viscoelastic stress. Focusing on parameters close to a Hopf bifurcation, where the uncontrolled system displays oscillatory states as well as hysteresis in the shear rate controlled protocol, we demonstrate that (local) Pyragas control leads to a full stabilization to the steady state solution of the total stress, while a global control scheme does not work. In contrast, for the case of imposed total stress, global Pyragas control fully stabilizes the system. In both cases, the control does not change the space of solutions, rather it selects the steady state solutions out of the existing solutions. This underlines the non-invasive character of the Pyragas scheme.%
\end{abstract}

\pacs{47.57.Qk, 05.45.Gg, 47.52.+j, 02.30.Yy}

\maketitle


\section{Introduction}\label{sec:I}

Exposing soft matter systems to shear flow often results in a non-linear behavior of measurable rheological quantities (such as viscosity, stress) and to instabilities. The most prominent example is the shear banding instability, where the system splits into two or multiple spatial regions (``bands'') of different local shear rate and local viscosity \cite{Olmsted2000,Fielding2007,Olmsted2008,Divoux2016}. This instability is indicated by a non-monotonicity of the flow curve, that is, the relationship between the total stress $\Ttot$ in the steady state and the imposed shear rate $\dot{\gamma}$. Shear banding occurs in many complex fluids like polymer solutions, emulsions or liquid crystals. To capture this phenomenon theoretically, several non-linear models have been proposed revealing multiple branches of the underlying constitutive curve $\Ttot(\dot{\gamma})$. A classical example is the diffusive Johnson-Segalman model \cite{Johnson1977,Olmsted2000}.%

In the present article we concentrate on one of the most prominent examples of a shear-banding system, namely solutions of elongated or wormlike micelles. Within the past decades the self-assembly processes as well as the rheological properties of such systems have attracted a lot of interest \cite{Berret2006,Cates2006,Dreiss2007,Lerouge2010}. Due to their flexibility, on the one hand, and their ability to break and reform, on the other hand, wormlike micelles often called ``living polymers''. Further, wormlike micellar solutions exhibit an unique viscoelastic behaviour which can be exploited for many applications such as daily care products \cite{Yang2002} or as fracturing fluids in oil recovery \cite{Sullivan2007}. Indeed, already the first rheological experiments done by Rehage and co-workers demonstrated that the viscosity of these elongated micelles reveals a highly non-linear flow behaviour \cite{Wunderlich1987,Rehage1988}: Unlike covalently bound polymers, wormlike micelles can not only break but also recombine after exposing them to high shear forces. The first theory to account for the resulting viscoelastic spectra of wormlike micelles has been proposed by Turner and Cates \cite{Cates1990,Turner1991}.%

Beyond stationary shear bands and nonlinear viscosity, wormlike micellar solutions can moreover display oscillatory states and irregular or even chaotic spatio-temporal patterns, as indicated based on a bunch of experimental data based on particle image velocimetry and bulk-rheology \cite{Salmon2002,Ganapathy2006,Ganapathy2006a,Ganapathy2008}. Theoretical approaches to account for this complex behavior in micellar (and other complex) fluids may be subdivided into three classes. The first are spatially homogeneous models for the rheological (mechanical) dynamical variables such as $\Ttot$ (or the viscoelastic stress $\sigma$) and $\dot{\gamma}$; this approach has been applied to shear-thickening fluids \cite{Cates2002,Head2002}. The same variables are also at the center of several inhomogeneous or non-local models for shear-thinning \cite{Fielding2003,Fielding2004} and shear-thickening fluids \cite{Aradian2005}. These models additionally take into account that the macroscopic rheological behaviour is coupled to a intrinsic (microscopic) variable such as the polymer concentration \cite{Fielding2003}, the micellar length \cite{Fielding2004}, or the director of orientational ordering. The third class focuses directly on the (spatio-temporal) dynamics of liquid-crystalline substances, involving evolution equations for the (tensorial) nematic order parameter and its backcoupling to the flow \cite{Rienacker2002,Rienacker2002a,Chakrabarti2004,Das2005}.%

The present study starts from a (non-local) model for micellar solutions proposed by Fielding and Olmsted \cite{Fielding2004}, where the viscoelastic stress $\sigma(y,t)$ is coupled to the length $n(y,t)$ of the elongated micelles. This coupling of a mechanical and a structural variables destabilizes the high shear rate branch of the flow curve, giving rise to spatio-temporal oscillatory and chaotic solutions beyond a Hopf bifurcation \cite{Fielding2004}. The goal of the present paper is twofold: First, we systematically examine the system's dynamics by a linear stability analysis for two rheological protocols (fixed average $\dot\gamma$ or $\Ttot$), thereby extending the results presented in Ref.~\cite{Fielding2004}. Second, we explore methods to manipulate the occurring dynamical states by using time-delayed feedback control.%

Indeed, in many technical applications of wormlike micelles the occurrence of oscillatory or chaotic states is not desired. Motivated by this, we here seek to stabilize {\it stationary} states inside the parameter range where the uncontrolled system displays oscillations. Specifically, we apply the concept of time-delayed feedback control (TDFC) as proposed by Pyragas in 1992~\cite{Pyragas1992}. Within the Pyragas control scheme, the dynamical equations for the relevant variables $a$ (which, ideally, should be measurable quantities) are supplemented by control terms involving the difference $a(t)-a(t-\tau)$, with $\tau$ being the delay time. This type of control is {\em noninvasive} as the control forces vanish when the steady state (or a periodic state with period $T_\mathrm{osc}=m\tau$ with $m =1,2,\ldots$) is reached. The Pyragas scheme has been applied to a broad variety of non-linear systems from various fields (see \cite{Schoell2008,Schoell2016} for overviews) including semiconductor nanostructures, lasers, neural systems and general reaction-diffusion systems \cite{Kyrychko2009,Gurevich2013}. There are also recent applications to the flow of soft-matter systems, both from the experimental \cite{Luthje2001} and from the theoretical side \cite{Zeitz2015a}, an example being the stabilization of steady shear-aligned states in sheared liquid crystals \cite{Strehober2013}. Within the present study, we focus on a non-diagonal control scheme which targets the average total stress or shear rate, respectively, yielding a Pyragas term involving only the viscoelastic stress. We apply this scheme to states close to a Hopf bifurcation, where the system displays hysteresis and the relation between shear rate and total stress is not unique. The control then selects the stationary solution.%

Our publication is structured as follows. In Sec.~\ref{sec:II} we review the model equations proposed in Ref.~\cite{Fielding2004}. In Sec.~\ref{sec:III} we describe the linear stability analysis for the two rheological protocols and compare the results with those from extensive numerical calculations of the full system. In Sec.~\ref{sec:IV} we introduce a time-delayed feedback control scheme to both rheological protocols. We then use the results of linear stability analysis derived in Sec.~\ref{sec:III} to find the neutral stability curves determining the parameter ranges where the control scheme is successful. The article closes with a brief summary and outlook.

\section{Model Equations and Background}\label{sec:II}

\subsection{Flow geometry and force balance equation}\label{sec:II_A}

We assume our micellar system to be subject to a planar Couette geometry consisting of two plates separated by a distance $L$ along the $y$-direction. The lower plate is fixed at the position $y=0$, while the upper plate located at the distance $y=L$ moves with a constant velocity in $x$-direction. This is equivalent to an externally imposed velocity field $\velocity^0=v_x^0(y,t)\,\unitvector$, with $\unitvector$ being the unit vector in $x$-direction and $v_x^0(y,t)=y\,\overline{\dot{\gamma}}(t)$. Here, $\overline{\dot{\gamma}}(t)$ is the external shear rate corresponding to $\velocity^0$. The \textit{local} shear rate can be defined as spatial gradient of the actual velocity profile $v_x(y,t)$ in the system (which may differ from $v_x^0(y,t)$), that is, $\dot{\gamma}(y,t)=\partial_y\,v_x(y,t)$. In general, the total stress tensor $\Ttensor$ of a viscoelastic fluid is defined by \cite{Larson1999}%
\begin{equation}
\Ttensor=\VStensor+2\eta\,\Deform-p\,\Imat\:,
\label{eq:1}
\end{equation}

where $\VStensor$ is the viscoelastic contribution which may be related to a structual quantity. The second term on the r.h.s. of Eq.~(\ref{eq:1}) denotes the Newtonian part, with $\eta$ being the Newtonian viscosity and $2\Deform=\nabla\,\velocity+(\nabla\,\velocity)^T$ the deformation tensor. The third term in Eq.~(\ref{eq:1}) is an isotropic contribution, where $p$ is the hydrostatic pressure and $\Imat$ the identity matrix. 

The dynamics of a viscoelastic fluid can be described through the generalized Navier-Stokes equation (or momentum balance equation)%
\begin{equation}
\begin{aligned}
\rho\,\Big(\partial_t+\velocity\,\nabla\Big)\,\velocity & =\nabla\,\Ttensor\\
 & =\nabla\,\Big(\VStensor+2\eta\,\Deform-p\,\Imat\Big)\:,
\end{aligned}
\label{eq:2}
\end{equation}

where $\rho$ is the fluid density. We assume our system to be incompressible, i.e. $\nabla\cdot\velocity=0$. Additionally, the system is assumed to be overdamped (limit of low Reynolds numbers). In this case the momentum balance equation~(\ref{eq:2}) reduces to the force balance equation%
\begin{equation}
\nabla\,\Big(\VStensor+2\eta\,\Deform-p\,\Imat\Big)=0\:.
\label{eq:3}
\end{equation}

For the present imposed linear velocity field, Eq.~(\ref{eq:3}) reduces to%
\begin{equation}
\partial_y\,\Big(\sigma(y,t)+\eta\,\dot{\gamma}(y,t)\Big)=0\:,
\label{eq:4}
\end{equation}

where $\sigma(y,t)$ is the local value of the $xy$-component of the viscoelastic stress tensor $\VStensor$, and $\dot{\gamma}(y,t)$ is the local shear rate. The restriction to one spatial dimension is motivated by our flow geometry. The force balance equation (\ref{eq:4}) demands the $xy$-component $\Ttot(y,t)$ of the stress tensor $\Ttensor$ to be spatially constant. Thus
\begin{equation}
\Ttot(y,t)=\sigma(y,t)+\eta\,\dot{\gamma}(y,t)\equiv\Ttot(t)
\label{eq:5}
\end{equation}

for all positions $y$ and times $t$. Note that the constancy in space of $\Ttot(t)$ does not necessarily hold for the individual term $\sigma(y,t)$ and $\dot{\gamma}(y,t)$. Averaging the second and third part of Eq.~(\ref{eq:5}) over the $y$-coordinate yields
\begin{equation}
\overline{\sigma}(t)+\eta\,\overline{\dot{\gamma}}(t)=\Ttot(t)\:.
\label{eq:6}
\end{equation} 

where $\overline{A}(t)=\left\langle A(y,t)\right\rangle_y=1/L\,\int_0^L\,A(y,t)\,\mathrm{d}y$ defines the spatial average over the $y$-direction of a quantity $A(y,t)$, which can be either the viscoelastic stress or the shear rate.

In the present study we consider two rheological protocols: First, we impose a temporarily constant average shear rate $\overline{\dot{\gamma}}(t)=\overline{\dot{\gamma}}=const.$ to the system, yielding the local viscoelastic stress $\sigma(y,t)$ as the main dynamical variable. Equating the right hand sides of Eqs.~(\ref{eq:5}) and (\ref{eq:6}) yields an expression for the local shear rate $\dot{\gamma}(y,t)$, that is%
\begin{equation}
\dot{\gamma}(y,t)=\overline{\dot{\gamma}}+\dfrac{1}{\eta}\,\Big[\overline{\sigma}(t)-\sigma(y,t)\Big]\:.
\label{eq:7}
\end{equation}

The total stress in this shear rate controlled protocol is determined from Eq.~(\ref{eq:5}) or (\ref{eq:6}). Secondly, we impose a temporarily constant stress to the system, i.e., $\Ttot(t)=\Ttot^{\mathrm{fix}}=const$. From Eq.~(\ref{eq:1}) it then follows that%
\begin{equation}
\dot{\gamma}(y,t)=\dfrac{1}{\eta}\,\Big[\Ttot^{\mathrm{fix}}-\sigma(y,t)\Big]\:.
\label{eq:8}
\end{equation}

\subsection{Model equations in the shear rate controlled protocol}\label{sec:II_B}

In the present work we employ the rheological model proposed in Ref. \cite{Fielding2004}, which couples the viscoelastic stress $\sigma(y,t)$ to a variable characterizing the microscopic structure. The latter is given by the length distribution $n(y,t)$ of elongated micelles.%

The dynamics of $n(y,t)$ in the shear rate controlled protocol is governed by the non-linear relaxation equation%
\begin{equation}
\partial_t\,n(y,t)=-\dfrac{n(y,t)}{\tau_n}+\dfrac{1}{\tau_n}\,\left(\dfrac{n_0}{1+\left[\tau_n\,\dot{\gamma}(y,t)\right]^\beta}\right)\:,
\label{eq:9}
\end{equation}

where $\tau_n$ is the relaxation time of the underlying rates of scission and recombination of the micelles, $n_0>0$ is the (average) length at zero shear and $\beta\ge0$ is a power law exponent. While the first term on the right hand side of Eq.~(\ref{eq:9}) is a pure relaxation term, the second term accounts for shear-induced changes of scission or recombination processes. The local shear rate $\dot{\gamma}(y,t)$ is determined through Eq.~(\ref{eq:7}).%

The homogeneous steady state solution ($\partial_t\,n(y,t)=0$) of Eq.~(\ref{eq:9}) reads%
\begin{equation}
n_s(\overline{\dot{\gamma}})=\dfrac{n_0}{1+\left(\tau_n\,\overline{\dot{\gamma}}\right)^\beta}\:.
\label{eq:10}
\end{equation}

\begin{figure}
\centering
\includegraphics[scale=1]{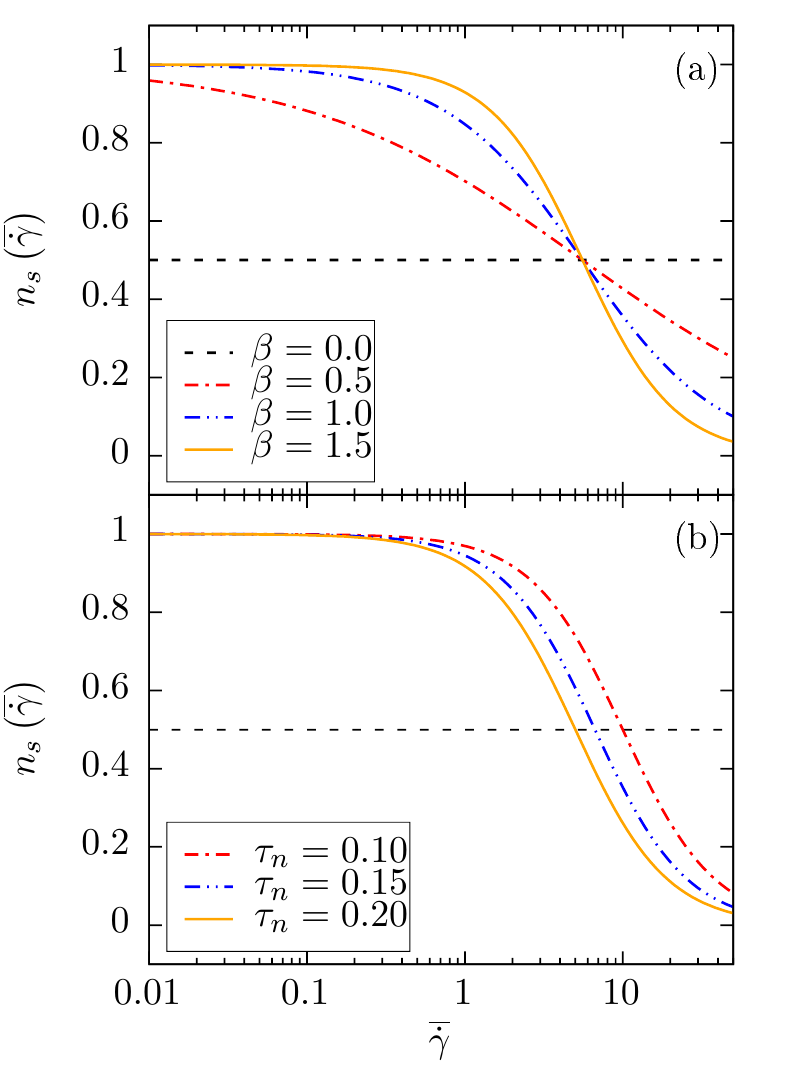}
\caption{Steady state solution $n_s$ as function of the imposed shear rate $\overline{\dot{\gamma}}$ for (a) varying power law exponents $\beta$ at $\tau_n=0.180$ and (b) varying relaxation times $\tau_n$ at $\beta=1.5$. The dashed horizontal line in (b) indicates where $n_s$ is halved.}
\label{fig:1}
\end{figure}

In the low shear rate limit, $\overline{\dot{\gamma}}\rightarrow0$, the steady state solution converges to $n_0$, while in the limit of (infinitely) high shear rates ($\overline{\dot{\gamma}}\rightarrow\infty$) we find $n_s\rightarrow0$. The rate of convergence in this high shear rate limit strongly depends on the power law exponent $\beta$ as illustrated in Fig. \ref{fig:1} $(a)$. In the case $\beta=0$ the steady state solution will be $n_s=n_0/2=const.$ for all shear rates. The dependence of the function $n_s(\overline{\dot{\gamma}})$ on the relaxation time $\tau_n$ is shown in Fig. \ref{fig:1} $(b)$. We further note that there is a critical shear rate $\overline{\dot{\gamma}}_c=\tau_n^{-1}$, at which the zero-shear length $n_0$ is halved upon increase of $\overline{\dot{\gamma}}$. Generally, the decrease of the micellar length with increasing shear rates reflects that scission processes become dominant. A similar behaviour has been observed in particle-based (Molecular Dynamics) computer simulations \cite{Kroger1996,Padding2008}. Experimentally, the micellar length is hardly accessible.%

The spatio-temporal dynamics of the viscoelastic stress $\sigma(y,t)$ is determined by the partial differential equation%
\begin{equation}
\begin{aligned}
\partial_t\,\sigma(y,t)= & -\dfrac{\sigma(y,t)}{\tau(n)}+\dfrac{\dot{\gamma}(y,t)}{1+\left[\tau(n)\,\dot{\gamma}(y,t)\right]^2}\\
 & +\mathcal{D}\,\partial_y^2\,\sigma(y,t)\:,
\end{aligned}
\label{eq:11}
\end{equation}

with $\mathcal{D}$ being the (stress-)diffusion constant and $\tau(n)$ a length dependent relaxation time [not to be confused with the time $\tau_n$ appearing in Eq.~(\ref{eq:9})]. The quantity $\tau(n)$ is defined as%
\begin{equation}
\tau(n)=\tau_0\,\left(\dfrac{n(y,t)}{n_0}\right)^\alpha\:.
\label{eq:12}
\end{equation}

Here, $\tau_0>0$ is the zero-shear relaxation time of the micelles, and $\alpha\ge0$ is another power law exponent. The system of Eqs.~(\ref{eq:9}), (\ref{eq:11}) and (\ref{eq:12}) is closed by Eq.~(\ref{eq:7}) for $\dot{\gamma}(y,t)$.

The homogeneous steady state solution of Eq.~(\ref{eq:11}) (i.e., $\partial_t\,\sigma(y,t)=0$) in the non-diffusive limit ($\mathcal{D}=0$) takes the following form:%
\begin{equation}
\sigma_s(\overline{\dot{\gamma}})=\dfrac{\tau(n_s)\,\overline{\dot{\gamma}}}{1+\left(\tau(n_s)\,\overline{\dot{\gamma}}\right)^2}\:.
\label{eq:13}
\end{equation}

Here we have to distinguish several limiting cases: For $\alpha=0$, the relaxation time reduces to a constant ($\tau(n_s)=\tau_0$), and therefore the coupling to the structural variable $n(y,t)$ vanishes. This case has been discussed, e.g., in Ref. \cite{Spenley1996}. The steady state solution $\sigma_s$ is then equivalent to the steady state solution found for the $xy$-component of the viscoelastic stress tensor in the Johnson-Segalman model \cite{Malkus1990}. For $0<\alpha<1$, we find $n_0^\alpha\,\tau(n_s)/\tau_0>n_s^\alpha$ for all shear rates. On the other hand, $n_0^\alpha\,\tau(n_s)/\tau_0<n_s^\alpha$ for $\alpha>1$ for all shear rates. Finally, in the special case $\alpha=1$, the relaxation time follows instantaneously the micellar length, that is $\tau(n_s)\propto n_s$. The plausibility of the ansatz made in Eq.~(\ref{eq:12}) is underlined by results from Molecular Dynamics simulations showing that the breaking time, which is similar to the relaxation time defined in Eq.~(\ref{eq:12}), follows a power law like behaviour \cite{Padding2004}.%

Using Eqs.~(\ref{eq:6}) and (\ref{eq:13}) we can state an explicit expression for the total stress in the homogeneous stationary limit,%
\begin{equation}
\Ttot_s(\overline{\dot{\gamma}})=\sigma_s(\overline{\dot{\gamma}})+\eta\,\overline{\dot{\gamma}}=\dfrac{\tau(n_s)\,\overline{\dot{\gamma}}}{1+\left(\tau(n_s)\,\overline{\dot{\gamma}}\right)^2}+\eta\,\overline{\dot{\gamma}}\:.
\label{eq:14}
\end{equation}

\begin{figure}
\centering
\includegraphics[scale=1]{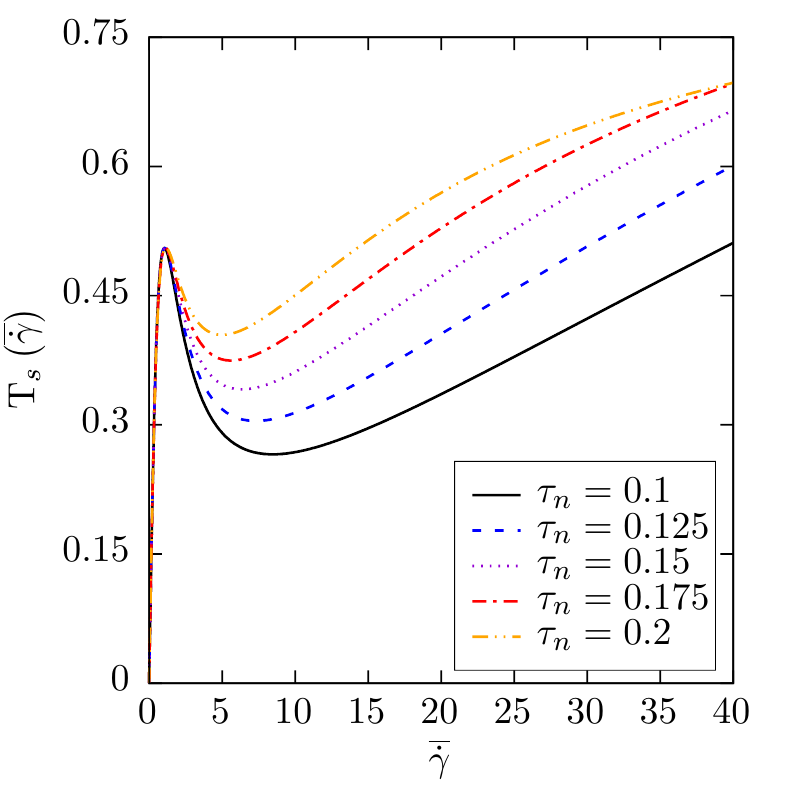}
\caption{Steady state solutions of the total stress for various relaxation times (fixed parameters: $\alpha=1.2$, $\beta=1.5$ and $\eta=0.005$).}
\label{fig:2}
\end{figure}

From the Johnson-Segalman model it is known that $\sigma_s$ as function of $\overline{\dot{\gamma}}$ displays a maximum at low shear rates and then converges towards zero for increasing shear rates. For the present model, the behaviour of $\sigma_s$ as function of $\overline{\dot{\gamma}}$ strongly depends on the choice of the power law exponent $\alpha$: For $\alpha<0.5$, we find the same behaviour as in the Johnson-Segalman model. However, for $\alpha\ge0.5$ this behaviour changes such that the maximum at low shear rate is followed by two other extrema, that is one minimum and one maximum, before tending to zero at high shear rates. This non-monotonic behaviour reflects the highly non-linear character of the model equations~(\ref{eq:9}) and (\ref{eq:11}). To get a more detailed insight into the impact of changing the relaxation time $\tau_n$, which is attributed to scission and recombination processes, we plot in Fig. \ref{fig:2} the steady state solution of the total stress, $\Ttot_s(\overline{\dot{\gamma}})$, for various $\tau_n$. Upon increase of $\tau_n$ the maximum of the constitutive curve stays at the same position, while the minimum of the curve shifts to smaller shear rates. A further effect of increasing $\tau_n$ is a change of the slope of the high shear rate branch. This is again an indication for a stronger coupling to the (micro-)structure.%

\subsection{Model equations in the stress controlled protocol}\label{sec:II_C}

The dynamical equations corresponding to the stress controlled protocol ($\Ttot(t)=\Ttot^{\mathrm{fix}}=const.$) follow from Eqs.~(\ref{eq:9}) and (\ref{eq:11}) by replacing the shear rate via Eq.~(\ref{eq:8}), that is, $\dot{\gamma}(y,t)=1/\eta\,[\Ttot^{\mathrm{fix}}-\sigma(y,t)]$. This leads to%
\begin{equation}
\begin{aligned}
\partial_t\,n(y,t)= & -\dfrac{n(y,t)}{\tau_n}\\
 & +\dfrac{1}{\tau_n}\,\left(\dfrac{n_0}{1+\Big[\tau_n\,\eta^{-1}\,\left(\Ttot^{\mathrm{fix}}-\sigma(y,t)\right)\Big]^\beta}\right)
\end{aligned}
\label{eq:15}
\end{equation}

and%
\begin{equation}
\begin{aligned}
\partial_t\,\sigma(y,t)= & -\dfrac{\sigma(y,t)}{\tau(n)}+\dfrac{\left(\Ttot^{\mathrm{fix}}-\sigma(y,t)\right)}{\eta+\Big[\tau(n)\cdot\left(\Ttot^{\mathrm{fix}}-\sigma(y,t)\right)\Big]^2}\\
 & +\mathcal{D}\,\partial_y^2\,\sigma(y,t)\:,
\end{aligned}
\label{eq:16}
\end{equation}

with%
\begin{equation}
\tau\left(n\right)=\tau_0\,\left(\dfrac{n(y,t)}{n_0}\right)^\alpha\:. 
\label{eq:17}
\end{equation}

The system of equations (\ref{eq:15}) to (\ref{eq:17}) is closed, i.e., there is no further equation required to calculate the dynamics of the system. 

The stress controlled protocol has already been discussed in Ref. \cite{Fielding2004} for the non-diffusive limit ($\mathcal{D}=0$), where the dynamical variables are only time-dependent. To evaluate the impact of a finite diffusion constant ($\mathcal{D}>0$) within the stress controlled protocol, we have performed numerical test calculations of Eqs.~(\ref{eq:15}) -- (\ref{eq:17}) using the procedure described in Sec. \ref{sec:III} (which allows for inhomogeneous solutions). We found the system to be spatially constant in all cases considered. Therefore, we set $\mathcal{D}=0$ in all following calculations related to the stress controlled protocol, yielding a spatially homogeneous dynamics.

\section{Linear Stability Analysis and State Diagrams} \label{sec:III}

In order to get a first insight into the (complex) dynamics of our model system, we perform a linear stability analysis. We follow the strategy outlined in Ref. \cite{Moorcroft2014}. We then compare the results with those from a full numerical solution. As we will demonstrate, the overall dynamics strongly depends on the rheological protocol. We therefore discuss the two protocols separately.%

\subsection{Shear rate controlled protocol} \label{sec:III_A}

\subsubsection{Linear stability analysis} \label{sec:III_A_1}

We start by introducing a two-dimensional vector containing the dynamical variables $n(y,t)$ and $\sigma(y,t)$,%
\begin{equation}
\vektor{\xi}(y,t)=\Big[n(y,t),\sigma(y,t)\Big]^T\:.
\label{eq:18}
\end{equation}

Within this vector notation we can rewrite Eqs.~(\ref{eq:9}) and (\ref{eq:11}) as%
\begin{equation}
\partial_t\,\vektor{\xi}(y,t)=\vektor{Q}\left(\vektor{\xi},\dot{\gamma}\right)+\partial_y^2\,\left(\Tensor{\mathcal{D}}\,\vektor{\xi}(y,t)\right)\:,
\label{eq:19}
\end{equation}

where $\vektor{Q}\left(\vektor{\xi},\dot{\gamma}\right)$ contains the non-linear terms on the r.h.s. of Eqs.~(\ref{eq:9}) and (\ref{eq:11}). Further, we have introduced the diffusion matrix $\Tensor{\mathcal{D}}$ which is given by%
\begin{equation}
\Tensor{\mathcal{D}}=\begin{pmatrix}0 & 0\\ 0 & \mathcal{D}\end{pmatrix}\:.
\label{eq:20}
\end{equation}

The homogeneous steady state solutions of the dynamical system Eq.~(\ref{eq:19}) are given in Eqs.~(\ref{eq:10}) and (\ref{eq:13}), respectively. We summarize these solutions in a vector $\vektor{\xi}_s=\left[n_s,\sigma_s\right]^T$. To determine whether $\vektor{\xi}_s$ is stable or unstable, we add small heterogeneous perturbations $\delta\vektor{\xi}(y,t)$, yielding%
\begin{equation}
\vektor{\xi}(y,t)=\vektor{\xi}_s+\delta\vektor{\xi}(y,t)\:.
\label{eq:21}
\end{equation}

Due to the coupling between shear rate and viscoelastic stress given in Eq.~(\ref{eq:7}), we additionally have to consider heterogeneous perturbations $\delta\dot{\gamma}(y,t)$ in the shear rate%
\begin{equation}
\dot{\gamma}(y,t)=\overline{\dot{\gamma}}+\delta\dot{\gamma}(y,t)\:.
\label{eq:22}
\end{equation}

Substituting Eqs.~(\ref{eq:21}) and (\ref{eq:22}) into Eq.~(\ref{eq:19}) and linearising up to first order in the perturbations leads to%
\begin{equation}
\begin{aligned}
\partial_t\,\delta\vektor{\xi}(y,t)= & \Mmat\cdot\delta\vektor{\xi}(y,t)+\vektor{q}\cdot\delta\dot{\gamma}(y,t)\\
 & +\partial_y^2\,\Big(\Tensor{\mathcal{D}}\cdot\delta\vektor{\xi}(y,t)\Big)\:,
\end{aligned}
\label{eq:23}
\end{equation}

where $\Mmat=\left.\partial_{\vektor{\xi}}\,\vektor{Q}\right|_{\vektor{\xi}_s,\overline{\dot{\gamma}}}$ and $\vektor{q}=\left.\partial_{\dot{\gamma}}\,\vektor{Q}\right|_{\vektor{\xi}_s,\overline{\dot{\gamma}}}$. To eliminate the perturbations in the shear rate we use the force balance equation~(\ref{eq:4}), which in its linearised form is given by%
\begin{equation}
0=\vektor{p}\cdot\delta\vektor{\xi}(y,t)+\eta\,\delta\dot{\gamma}(y,t)\:.
\label{eq:24}
\end{equation}

Here $\vektor{p}=(0,1)$ is a projection vector. Combining Eqs.~(\ref{eq:22}) and (\ref{eq:24}) then yields%
\begin{equation}
\partial_t\,\delta\vektor{\xi}(y,t)=\Big(\Jacobian+\partial_y^2\,\Tensor{\mathcal{D}}\Big)\,\delta\vektor{\xi}(y,t)\:,
\label{eq:25}
\end{equation}

where $\Jacobian$ is the Jacobian matrix given by%
\begin{equation}
\Jacobian=\Mmat-\dfrac{1}{\eta}\,\vektor{q}\cdot\vektor{p}\:.
\label{eq:26}
\end{equation}

Explicit expressions for the elements of $\Jacobian$ are given in Appendix \ref{Appendix_A}. We now assume the heterogeneous perturbations to be of exponential form such that%
\begin{equation}
\delta\vektor{\xi}(y,t)=\vektor{C}\,e^{\lambda t+iky}\:,
\label{eq:27}
\end{equation}

where $\vektor{C}=\left(C_1,C_2\right)^T$, $C_1,C_2\in\mathbb{R}$, is an arbitrary constant vector, $k$ is the wave number and $\lambda=\mu+i\omega$ is a complex number with $\mu,\omega\in\mathbb{R}$. The homogeneous steady state $\vektor{\xi}_s$ is said to be stable, if $\mu<0$, i.e., the perturbation dies out with time. Otherwise it is unstable. Inserting Eq.~(\ref{eq:27}) into Eq.~(\ref{eq:25}) leads to the eigenvalue problem%
\begin{equation}
\mathrm{det}\Big[\Jacobian-k^2\,\Tensor{\mathcal{D}}-\lambda\,\Imat\Big]=0\:.
\label{eq:28}
\end{equation}

\begin{figure}
\centering
\includegraphics[scale=1]{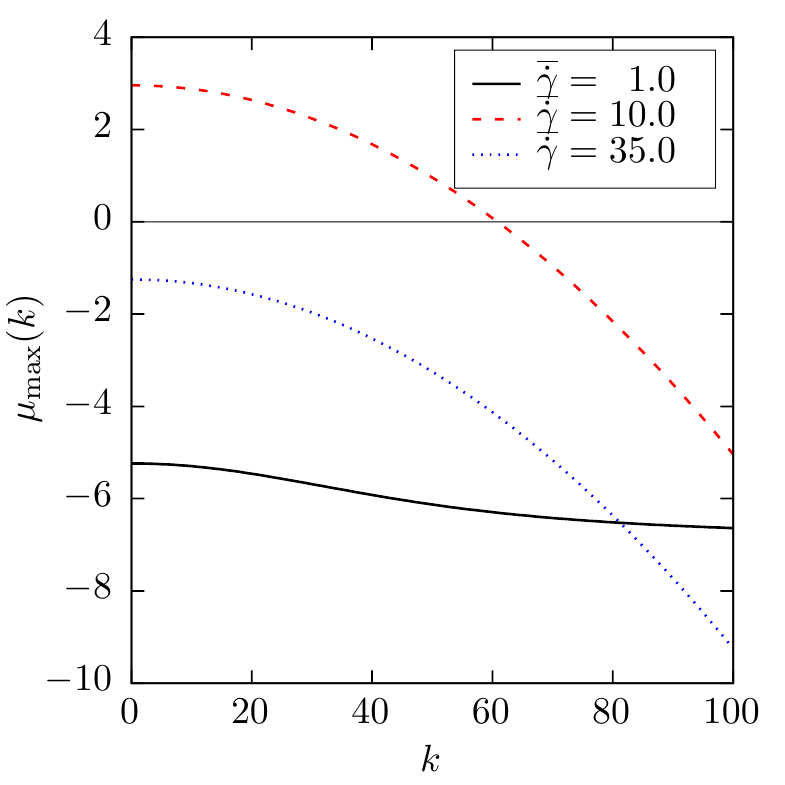}
\caption{Largest real part $\mu_{\max}$ of the complex eigenvalue $\lambda$ as function of the wave number $k$ for various imposed shear rates $\overline{\dot{\gamma}}$ at fixed $\tau_n=0.145$ and $\eta=0.005$.}
\label{fig:3}
\end{figure}

From Eq.~(\ref{eq:28}) it is seen that the complex eigenvalues $\lambda$ depends on the wave number $k$ and also on the imposed shear rate $\overline{\dot{\gamma}}$ (via $\Jacobian$). To see the impact of $k$ we plot in Fig. \ref{fig:3} the largest real part $\mu_{\max}=\mu_{\max}(k)$ (out of the two real parts of the two eigenvalues) at a fixed structural relaxation time $\tau_n$ and a fixed Newtonian viscosity $\eta$ for three imposed shear rates $\overline{\dot{\gamma}}$. In all cases, $\mu_{\max}(k)$ reaches its largest value at $k=0$. For $k>0$ the real part is a monotonic decaying function. This behaviour is a consequence of the asymmetric structure of Eq.~(\ref{eq:19}) regarding the diffusion terms. As a result, the eigenvalue equation~(\ref{eq:28}) contains only quadratic powers of $k$. The behaviour of $\mu_{\max}(k)$ indicates that the most unstable mode corresponds to $k=0$. In the following we therefore restrict our attention to this mode alone. Physically, this restriction implies to consider the system as homogeneous, in other words, $\mathcal{D}=0$.%

\begin{table}
\begin{tabular}{cc|ll}
real part & imag. part & character & \\
\hline
$\mu_{1,2}<0$ & $\omega_{1,2}=0$ & stable fix point & (sFP)\\
$\mu_{1,2}>0$ & $\omega_{1,2}=0$ & unstable fix point & (uFP)\\
$\mu_1>0$, $\mu_2<0$ & $\omega_{1,2}=0$ & unstable saddle & (uSAD)\\
$\mu_{1,2}<0$ & $\omega_1>0$, $\omega_2<0$ & stable focus & (sFOC)\\
$\mu_{1,2}>0$ & $\omega_1>0$, $\omega_2<0$ & unstable focus & (uFOC)
\end{tabular}
\caption{Classification of the complex eigenvalues $\lambda=\mu+i\omega$ in the special case of a homogeneous system.}
\label{tab:1}
\end{table}

\begin{figure}[htb]
\centering
\includegraphics[scale=1]{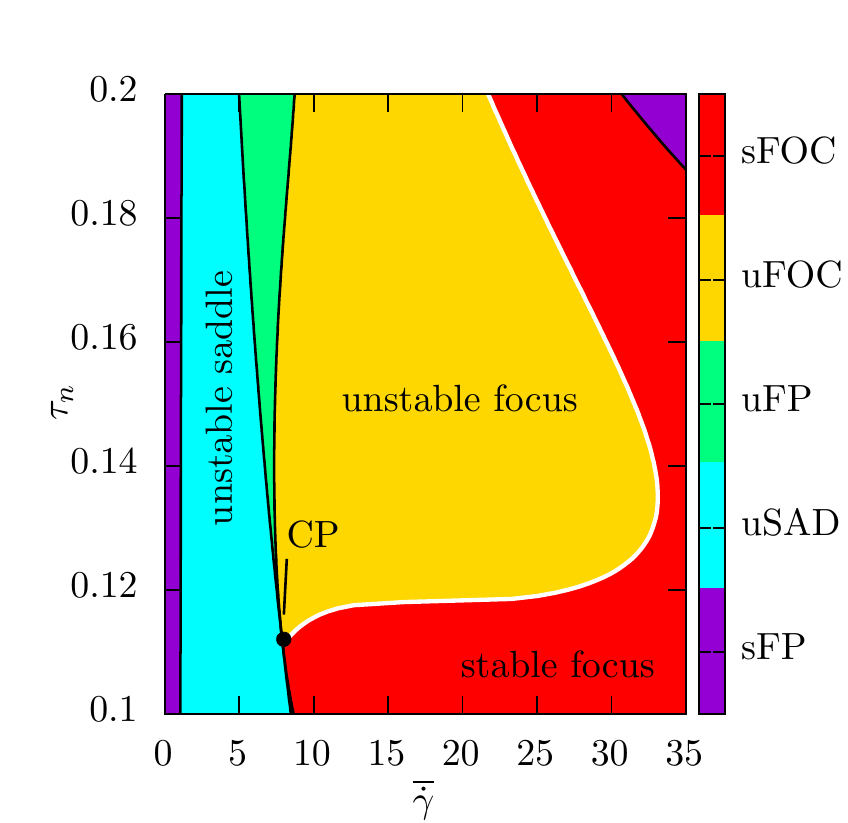}
\caption{Stability diagram obtained by solving the eigenvalue equation (\ref{eq:28}) for $k=0$ in a range of relaxation times $\tau_n$ and imposed shear rates $\overline{\dot{\gamma}}$. The states are classified according to table \ref{tab:1}. The Hopf bifurcation (solid white line) delimits the unstable and stale focus.}
\label{fig:4}
\end{figure}

The various stability scenarios occurring in this case can be classified according to table \ref{tab:1}. Investigating the complex eigenvalues for a range of (average) shear rates $\overline{\dot{\gamma}}$ and relaxation times $\tau_n$ we obtain a stability diagram, which is presented in Fig. \ref{fig:4}.%

Below a critical point (CP) \cite{critpoint} the system shows essentially the same behaviour as the Johnson-Segalman model: There are two stable regions (stable fix point and stable focus), corresponding to homogeneous steady states, and one unstable region (unstable saddle). Within the Johnson-Segalman model the occurrence of the unstable saddle is associated with the regime, where the slope of the constitutive curve $\Ttot_s(\overline{\dot{\gamma}})$ is negative, respectively where shear banding can occur \cite{Espanol1996}. Above the critical point the present system displays additionally an unstable fix point and an unstable focus. The latter is separated from the stable focus by a Hopf bifurcation (corresponding to $\mu_{1,2}=0$ and $\omega_1>0$, $\omega_2<0$). The Hopf bifurcation is absent in the Johnson-Segalman model and may thus be considered as a characteristic feature of the present model.%

\subsubsection{Numerical results} \label{sec:III_A_2}

Our stability analysis at $k=0$ clearly does not provide complete information on the system. To explore the full spatio-temporal dynamics at $\mathcal{D}\ne0$ we solved Eqs.~(\ref{eq:9}) and (\ref{eq:11}) numerically by using a Crank-Nicolson algorithm \cite{Press2002}. We sample over $N_t=30,000$ temporal and $N_y=150$ spatial discrete steps with a corresponding step size $dt=0.005$ and $dy=0.007$, yielding a total integration time of $T=N_t\cdot dt=150$, and a gap size of $L=1.0$. In analogy to Ref. \cite{Fielding2004}, we fixed the two power law exponents at $\alpha=1.2$ and $\beta=1.5$, and the diffusion constant at $\mathcal{D}=0.0016$. The equilibrium parameters, namely the zero-shear length and relaxation time, are set to $n_0=\tau_0=1.0$, respectively. To ensure numerical stability we use the Courant-Friedrichs-Lewy criterion for forward time stepping given through $\sqrt{2\,\mathcal{D}\,dt}<dy$ \cite{Courant1928}. To realize unique initial conditions for all numerical calculations, we prepared the stress to be homogeneous in space and the micellar length to be in a shear banded state given by $n(y,t=0)=\overline{N}\cdot\left[1+\Delta\cdot\cos\left(y\,\pi\right)\right]$. In order to check the sensitivity of the system with respect to the initial conditions, we varied the constants $\overline{N}$ and $\Delta$. The results presented here have been obtained with $\overline{N}=0.5$ and $\Delta=0.5$ corresponding to $\overline{n}(t=0)=n_0/2$. At the confining walls we employ Dirichlet boundary conditions through $\left.\partial_y\,\sigma(y,t)\right|_{y=0}=0$ and $\left.\partial_y\,\sigma(y,t)\right|_{y=L}=0$.%

\begin{figure*}[htb]
\centering
\includegraphics[scale=0.2]{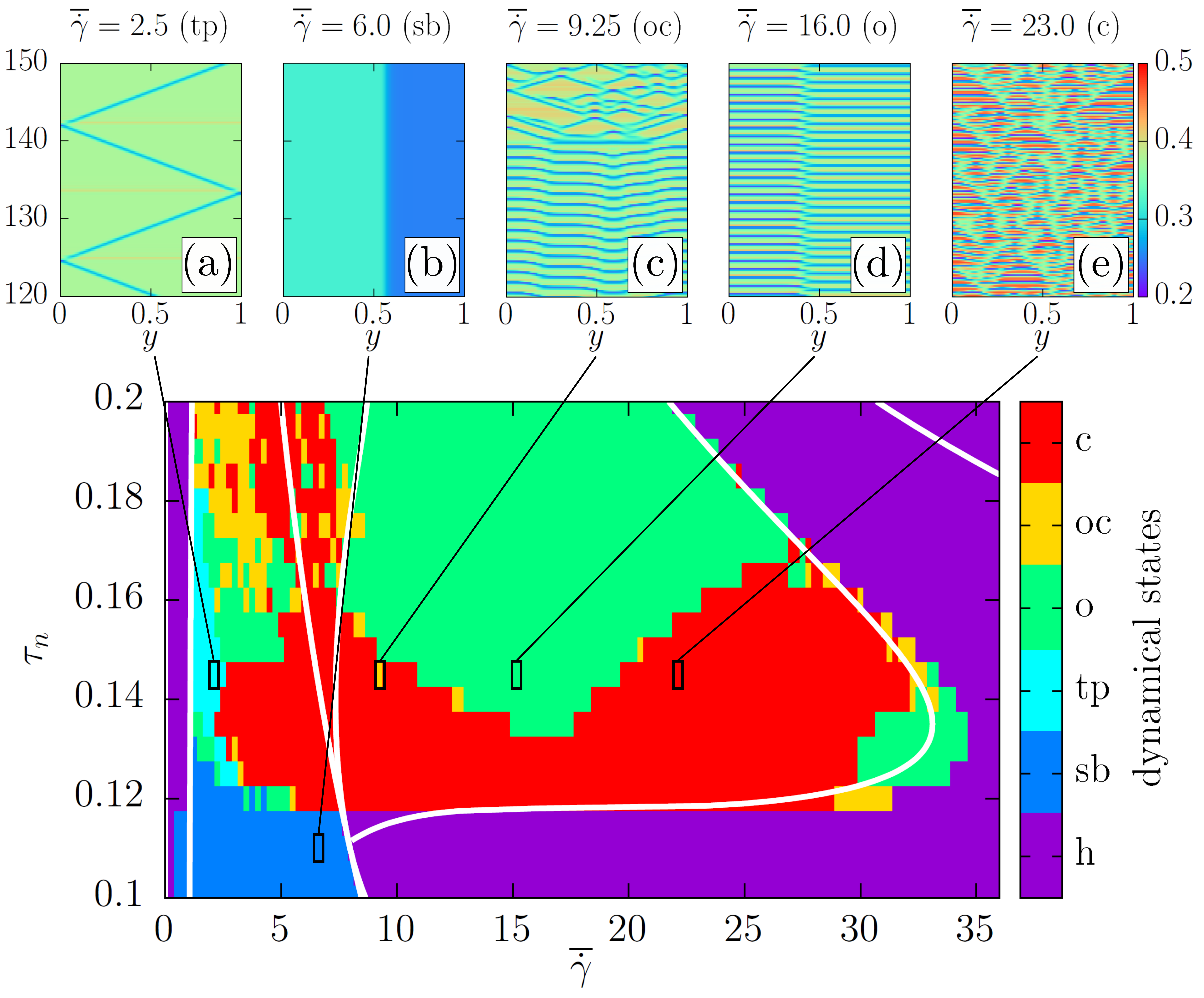}
\caption{Numerical results obtained at $\eta=0.005$ and $\mathcal{D}=0.0016$: The upper panel shows spatio-temporal patterns of the viscoelastic stress $\sigma(y,t)$ at different shear rates. These patterns correspond to (a) travelling pulses [tp], (b) shear banding [sb], (c) oscillatory-chaotic [oc], (d) oscillatory [o] and (e) chaotic [c] states. The lower panel shows the state diagram. The white lines result from the linear stability analysis. }
\label{fig:5}
\end{figure*}

In our numerical calculations we performed ``continuous shear ramps''. That is, starting from a given configuration of the dynamical variables, we solved the equations at constant $\overline{\dot{\gamma}}$ up to the steady state limit, which is typically achieved at $T=150$. To proceed towards the next (larger or smaller) value of the shear rate, we took the steady state configuration from the previous one as an initial configuration. By this strategy we are able to avoid stress ``overshoots'' in the start up regime of each shear step, leading to more stable numerical results. Further, this procedure is more realistic from an experimental point of view.%

In Fig. \ref{fig:5} we present the state diagram for a `down' ramp, that is starting from high shear rates and decreasing the shear rate step by step. It is seen that the system displays a variety of dynamical states, which differ from each other by the spatio-temporal behaviour of the viscoelastic stress $\sigma(y,t)$ and the other dynamical variables, that is, $n(y,t)$ and $\dot{\gamma}(y,t)$. We can distinguish six types of patterns (illustrated in Fig. \ref{fig:5}): spatially homogeneous (h), (static) shear banding (sb), travelling pulses (tp), spatio-temporal oscillatory (o), mixture of spatio-temporal oscillatory and chaotic (oc) and purely chaotic in space and time (c). Some of the results have already been reported in Ref. \cite{Fielding2004} and in fact, our results are consistent with this earlier study. Here we have substantially extended the range of relaxation times $\tau_n$ and imposed shear rates $\overline{\dot{\gamma}}$. In order to verify our classification made for the dynamical states we also determined the largest Lyapunov exponent.%

Inspecting the numerical results in Fig. \ref{fig:5} for small relaxation times $\tau_n\le0.112$, we see that the system behaves as predicted by linear stability analysis (and consistent with the Johnson-Segalman model). The regimes at low and high shear rates corresponding to homogeneous steady states are separated by a shear banded region associated with the unstable saddle. By increasing the relaxation time towards values above the critical point ($\tau_n>0.112$), where the Hopf bifurcation arises, we enter a region rich of dynamical states, which have already been observed by Fielding and Olmsted \cite{Fielding2004}. In the subsequent discussion we especially focus on the purely oscillatory and chaotic states (as identified by the Lyapunov exponents), respectively.%

\begin{figure}
\centering
\includegraphics[scale=1]{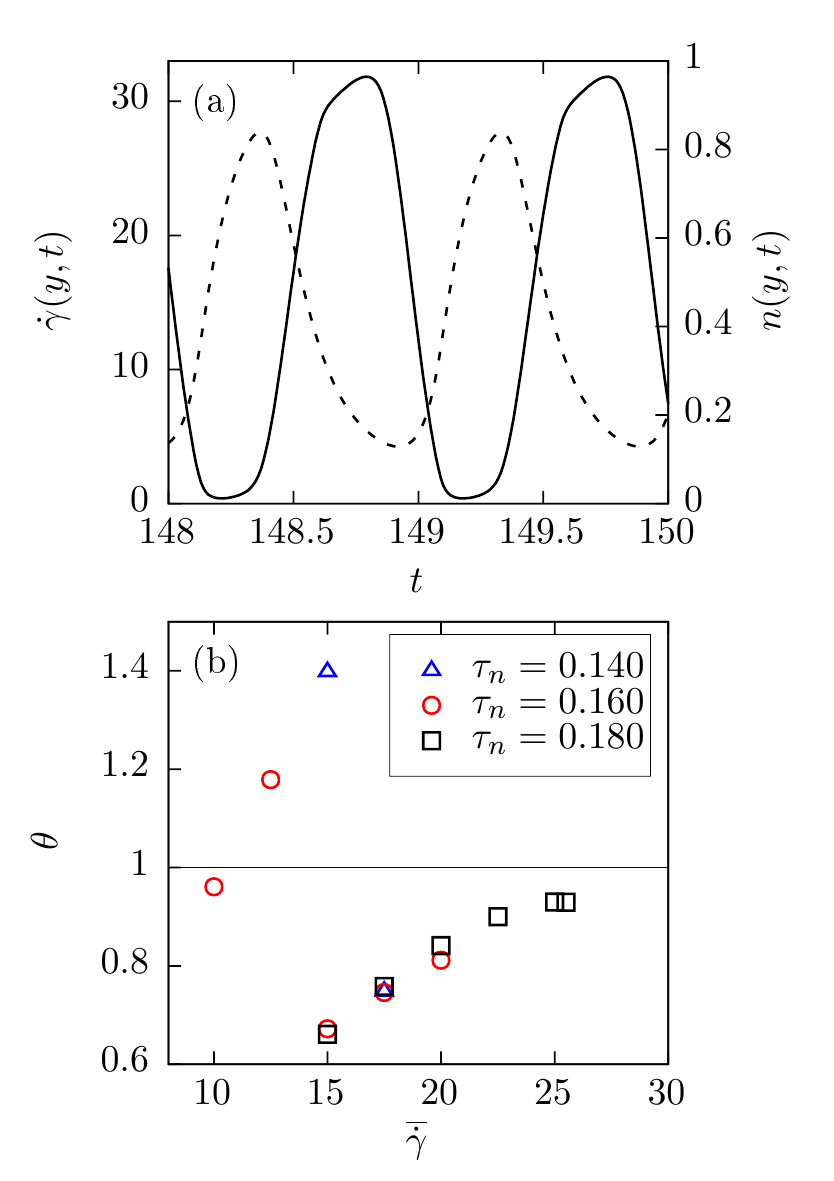}
\caption{(a) Temporal evolution of the local shear rate $\dot{\gamma}(y,t)$ (solid line) and the micellar length $n(y,t)$ (dashed line) at $y=0.8$ for $\overline{\dot{\gamma}}=16.0$ and $\tau_n=0.145$. (b) Frequency ratio $\theta$ for the oscillatory states within the unstable focus determined from power spectra at three relaxation times $\tau_n$ (fixed parameters: $\eta=0.005$, $\mathcal{D}=0.0016$, $\alpha=1.2$ and $\beta=1.5$).}
\label{fig:6}
\end{figure}

Within the oscillatory state the system consists of a well defined high and low shear rate band, which change position periodically in time. Similar states have been observed experimentally \cite{Azzouzi2005,Lerouge2008,Fardin2012}. In the present model the resulting oscillations of $\dot{\gamma}(y,t)$ are associated to oscillations of the micellar length $n(y,t)$. The time evolution of these quantities over one period as plotted in Fig. \ref{fig:6} (a) can be subdivided into four stages:%

\begin{itemize}
	\item[(i)] At very low local shear rate, the micellar length is maximal. From a physical point of view this corresponds to a situation, where the system consists of isotropically distributed and entangled micelles.
	\item[(ii)] The local shear rate starts to increase, while the micellar length decreases: This can be interpreted such that the micellar network starts to break down (due to increasing shear forces), while the micelles align in the shear flow.
	\item[(iii)] The local shear rate reaches its maximum, whereas the micellar length is found to be minimal: In this case the shear forces are so large that scission processes are dominating, yielding particularly small values of the micellar length.
	\item[(iv)] At the end of the cycle the local shear rate is still quite high but so is the degree of alignment of the micelles. Therefore, end-end-recombination processes are more likely to happen, and the micelles start to grow again.
\end{itemize}

To quantify the oscillatory dynamics inside the unstable regime we extract the fundamental frequency $f^\ast$ from a time series of the total stress $\Ttot(t)$ [see Eq.~(\ref{eq:6})] by creating a power spectrum. We then compare $f^\ast$ to the frequency $f=\omega/(2\pi)$ obtained from linear stability analysis by introducing the ratio $\theta=f/f^\ast$. Numerical results for $\theta$ are shown in Fig. \ref{fig:6} (b).

\begin{figure}
\centering
\includegraphics[scale=1]{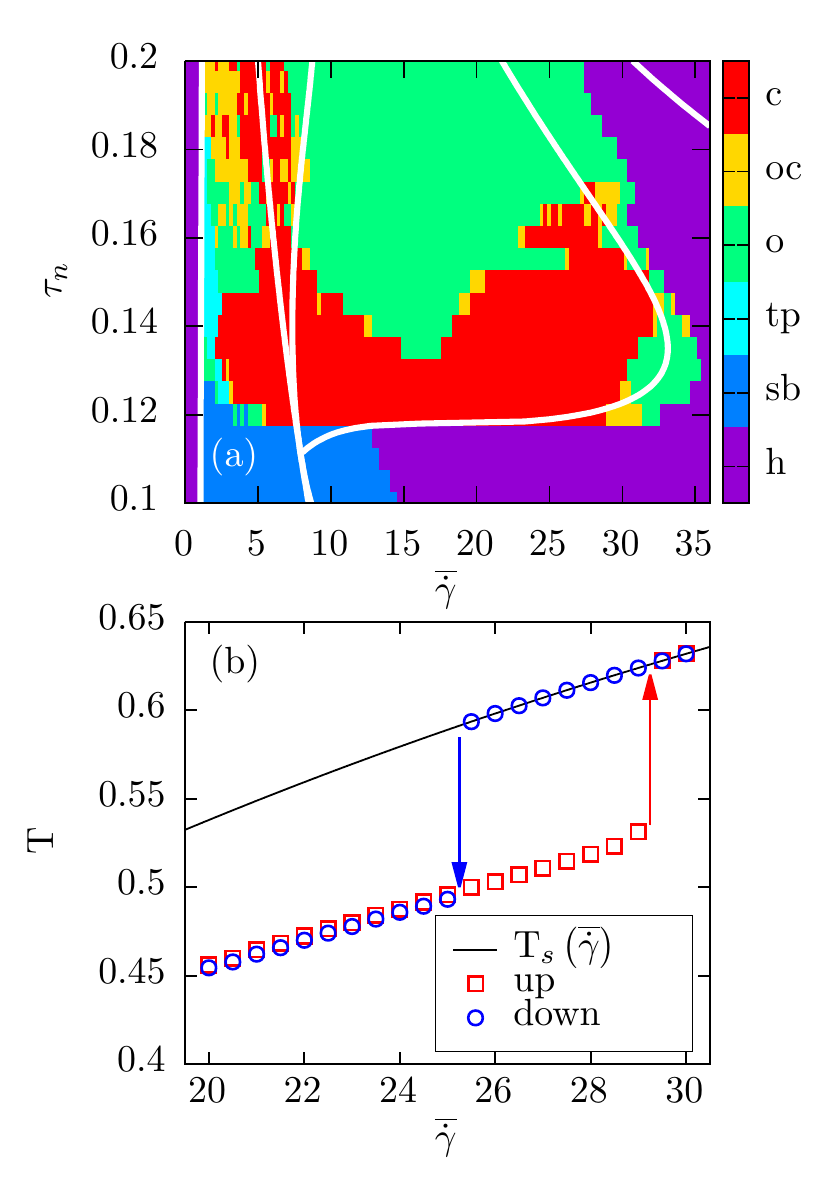}
\caption{(a) State diagram for the `up' ramp revealing the same dynamical states as classified in Fig. \ref{fig:5}. (b) Total stress $\Ttot$ as function of the imposed shear rate $\overline{\dot{\gamma}}$ at $\tau_n=0.180$ as given by the analytical steady state solution (solid line, Eq.~(\ref{eq:14})) and the numerical data for the up and down ramp (fixed parameters: $\eta=0.005$, $\mathcal{D}=0.0016$, $\alpha=1.2$ and $\beta=1.5$).}
\label{fig:7}
\end{figure}

Inspecting the frequency ratio $\theta$ for the oscillatory and chaotic states, we find that for low and moderate relaxation times, i.e. $\tau_n=0.140$ and $\tau_n=0.160$, no clear trend is visible. For high relaxation times, here $\tau_n=0.180$, the system reveals only oscillatory states. Generally, we find the linear stability results underestimates the frequency for the total investigated shear rate range. Especially at the lower bound of the unstable focus the numerically determined frequency is about a factor $2.5$ higher than the estimated one. We recall, however, that the linear stability analysis is restricted to homogeneous states, therefore, deviations are expected. Interestingly, close to the Hopf bifurcation the numerical and analytical results match quite well. These results already give hints where it is possible to manipulate the dynamics of the system using time-delayed feedback control.

So far we have been focusing on the `down' ramp. The corresponding state diagram for an `up' ramp (i.e. increasing $\overline{\dot{\gamma}}$ from low values) is presented in Fig. \ref{fig:7} (a). Comparing the two state diagrams in Figs. \ref{fig:5} and \ref{fig:7} (a) at low relaxation times, i.e. below the critical point (where only shear banded states exist) we conclude that the system reveals the same hysteric behaviour, which was already described for the Johnson-Segalman model elsewhere \cite{Adams2008}. However, comparing the state diagrams above the critical point, we find for the up ramp the oscillatory region to be extended towards higher shear rates. To clarify this point, we plot in Fig. \ref{fig:7} (b) numerical steady state results for the total stress for the two ramp protocols. While for the `down' ramp the numerical data leave the analytical steady state solution [see Eq.~(\ref{eq:14})] exactly at the Hopf bifurcation, the corresponding data for the `up' protocol remains below the analytical one even after crossing the predicted Hopf bifurcation. This indicates that the system reveals a second instability, which is not covered by our linear stability analysis.

\subsection{Stress controlled protocol}\label{sec:III_B}

For the linear stability analysis of the stress controlled protocol we follow the same procedure outlined in the previous section. As argued in Sec. \ref{sec:II_C}, the stress controlled protocol leads to spatially homogeneous dynamics. We therefore set $\mathcal{D}=0$ and define the vector $\vektor{\xi}(t)=\left[n(t),\sigma(t)\right]^T$ containing the dynamical variables of the system. Within this vector notation Eqs. (\ref{eq:15}) and (\ref{eq:16}) can be written as%
\begin{equation}
\partial_t\,\vektor{\xi}(t)=\vektor{Q}\left(\vektor{\xi},\Ttot^{\mathrm{fix}}\right)\:,
\label{eq:29}
\end{equation}

where $\vektor{Q}\left(\vektor{\xi},\Ttot^{\mathrm{fix}}\right)$ summarizes the right hand sides of the two dynamical equations. We recall here that, unlike in the shear rate controlled protocol, the total stress is set to a constant. Further, we introduce the vector $\vektor{\xi}_s=\left[n_s,\sigma_s\right]^T$ containing the steady state solutions of the dynamical system [see Eq.~(\ref{eq:29})]. To determine the stability of these steady states, we add small time-dependent perturbations $\delta\vektor{\xi}(t)$ to $\vektor{\xi}_s$%
\begin{equation}
\vektor{\xi}(t)=\vektor{\xi}_s+\delta\vektor{\xi}(t)=\vektor{\xi}_s+\vektor{C}\,e^{\lambda t}\:,
\label{eq:30}
\end{equation}

with $\vektor{C}=(C_1,C_2)^T$ being a constant vector with $C_1,C_2\in\mathbb{R}$ and $\lambda=\mu+i\omega$, $\mu,\omega\in\mathbb{R}$, being the complex eigenvalue. Inserting this ansatz into Eq. (\ref{eq:29}) and linearising up to first order in the perturbations leads to the eigenvalue problem%
\begin{equation}
\mathrm{det}\Big[\Jacobian-\lambda\,\Imat\Big]=0\:.
\label{eq:31}
\end{equation}

\begin{figure}
\centering
\includegraphics[scale=1]{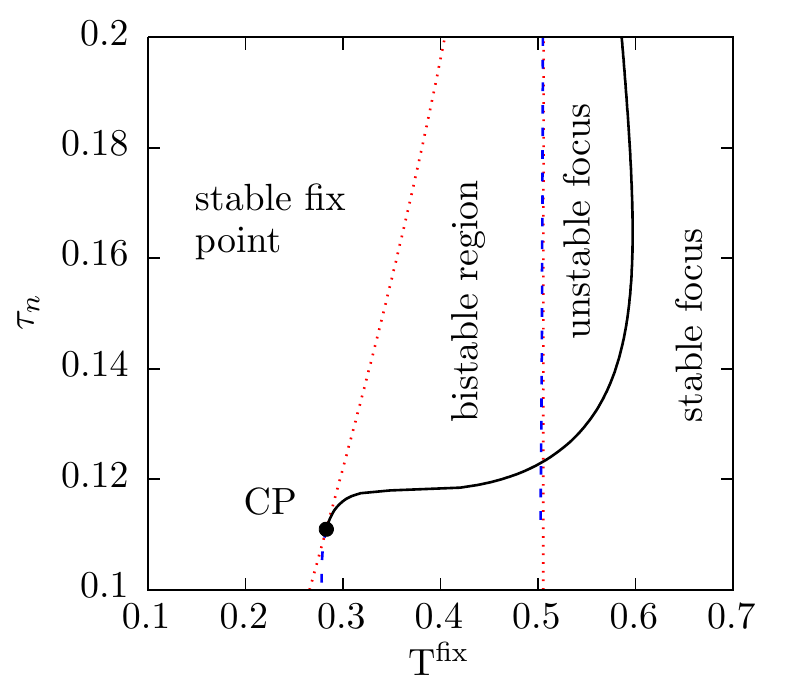}
\caption{Stability diagram for the stress controlled protocol obtained by solving the eigenvalue equation (\ref{eq:31}) in a range of relaxation times $\tau_n$ (fixed parameters: $\alpha=1.2$, $\beta=1.5$, $\eta=0.005$).}
\label{fig:8}
\end{figure}

Here $\Jacobian=\left.\partial_{\vektor{\xi}}\,\vektor{Q}\right|_{\vektor{\xi}_s,\Ttot^{\mathrm{fix}}}$ is the Jacobian matrix of the dynamical system, which is explicitly derived in Appendix \ref{Appendix_B}. When solving the eigenvalue equation (\ref{eq:31}) we find a quite simple stability diagram, see Fig. \ref{fig:8}, (compared to the shear rate controlled protocol). It reveals a stable fix point (homogeneous steady state), an unstable focus (homogeneous oscillatory state) delimited by a Hopf bifurcation, and a stable focus (homogeneous steady state). The transition between the region of stable fix points and unstable focus is characterised by hysteretic behaviour (bistability). This reduced complexity compared to Fig. \ref{fig:4} is a consequence of the fact that within the stress controlled protocol, there are only time-dependent states, such that the dimensionality of the system is reduced to $d=2$. This excludes to observe any chaotic states, since chaos requires dimension $d\ge3$ \cite{Strogatz1994}.

\section{Feedback Control}\label{sec:IV}

In this section we discuss the application of time-delayed feedback control (TDFC) to our model system. The primary goal thereby is to suppress oscillatory and chaotic states in favour of steady states. This is particularly motivated by an experimental study, where TDFC was used to control chaotic states appearing in Taylor-Couette flow of a complex fluid \cite{Luthje2001}. Generally, the dynamics of a shear-driven system can be manipulated by feeding back a measurable quantity such as the total stress $\Ttot(t)$ or the shear rate $\overline{\dot{\gamma}}(t)$ through a (closed) feedback loop. The control in our system is realized through a Pyragas scheme \cite{Pyragas1992}, where the equations of motion of the original system are supplemented by a control term of the form $a(t)-a(t-\tau)$. Here, $a$ is a measurable quantity, which can be either the total stress or the shear rate, at the present time $t$ and an earlier time $t-\tau$ with $\tau$ being the delay time. 

In the following we explore the impact of TDFC on both the shear rate and the stress controlled protocol. Thereby, we restrict ourselves to states close to the Hopf bifurcation, where the model system exhibits a hysteretic behaviour see, e.g.,  Fig. \ref{fig:7} for the shear rate controlled protocol. In this situation, the relation between (average) shear rate and stress is not unique. The TDFC proposed here aims at selecting one of the branches, particularly, the stationary branch of the two solutions.%

\subsection{Shear rate controlled protocol} \label{sec:IV_A}

The measurable quantity within the shear rate controlled protocol is the total stress $\Ttot(t)$ defined in Eq.~(\ref{eq:6}). Using this definition a Pyragas-like feedback term involving $\Ttot$ has the form%
\begin{equation}
\begin{aligned}
\Ttot(t)-\Ttot(t-\tau) & =\overline{\sigma}(t)+\eta\,\overline{\dot{\gamma}}(t)-\overline{\sigma}(t-\tau)-\eta\,\overline{\dot{\gamma}}(t-\tau)\\
 & =\overline{\sigma}(t)-\overline{\sigma}(t-\tau)\:.
\end{aligned}
\label{eq:32}
\end{equation}

Here, we assumed that the (mean) shear rate $\overline{\dot{\gamma}}(t)=\overline{\dot{\gamma}}(t-\tau)=\overline{\dot{\gamma}}=const.$ for all $t$. This seems somewhat non-physical due to the fact that, under normal circumstances, a control of the total stress should affect the shear rate (and vice versa). However, we demonstrated in Sec. \ref{sec:III_A_2} by numerically solving our (uncontrolled) model system, there are state points where the relationship between the total stress and the shear rate is not unique, rather one observes a hysteresis. In the following we will demonstrate that in such a situation, an application of the Pyragas feedback scheme leads to a ``stress selection'' in the system, that is for certain sets of control parameters $K$ and $\tau$, the system selects the steady state solution $\Ttot_s(\overline{\dot{\gamma}})$ given in Eq.~(\ref{eq:14}).%

Equation~(\ref{eq:32}) further implies that a feedback of the total stress $\Ttot$ is equivalent to a (global) feedback of the spatially averaged viscoelastic stress $\overline{\sigma}$. In other words, the TDFC affects only the dynamics of the viscoelastic stress and not that of the micellar length $n(y,t)$, yielding a non-diagonal control scheme. However, linear stability analysis suggests that the global TDFC cannot stabilize spatially inhomogeneous dynamics (see Appendix \ref{Appendix_C}), as confirmed by numerical test calculations.%

Therefore, we focus here on a local version of the control scheme, i.e. we use the local difference $\sigma(y,t)-\sigma(y,t-\tau)$ for all following calculations in Sec.~\ref{sec:IV_A}. We are aware that this type of control scheme is difficult to realize in an experimental set up, but from a theoretical point of view it is the most suitable way to introduce TDFC to the shear rate controlled protocol. However, we note already here that a global feedback scheme is successful within the stress controlled protocol (see Sec.~\ref{sec:IV_B}).%

In our vectorial notation introduced in Eq.~(\ref{eq:19}) for the uncontrolled system, the dynamical equation for the system under (local) TDFC is given by%
\begin{equation}
\begin{aligned}
\partial_t\,\vektor{\xi}(y,t)= & \:\vektor{Q}\left(\vektor{\xi},\dot{\gamma}\right)+\partial_y^2\,\left(\Tensor{\mathcal{D}}\,\vektor{\xi}(y,t)\right)\\
 & \:-K\,\Cmat\,\Big[\vektor{\xi}(y,t)-\vektor{\xi}(y,t-\tau)\Big]\:,
\end{aligned}
\label{eq:33}
\end{equation}

where $\Tensor{\mathcal{D}}$ is the diffusion matrix defined in Eq.~(\ref{eq:20}) and $\Cmat$ is a non-diagonal $2\times2$ matrix defined by%
\begin{equation}
\Cmat=\begin{pmatrix} 0 & 0\\ 0 & 1\end{pmatrix}\:.
\label{eq:34}
\end{equation}

This form ensures that the control is only applied to $\sigma(y,t)$. We note that a similar scheme has been proposed by Kyrychko and co-workers for the one-species control in the Gray-Scott model \cite{Kyrychko2009}, which belongs to the class of reaction-diffusion systems.%

We start again by deriving analytic expressions for the controlled eigenvalues $\nu=\mu+i\,\omega$ in the framework of linear stability analysis, following the procedure outlined for the uncontrolled case in Sec. \ref{sec:III_A_1}. Assuming heterogeneous perturbations $\delta\vektor{\xi}(y,t)$ and $\delta\dot{\gamma}(y,t)$, substituting these perturbations into Eq.~(\ref{eq:33}) and linearising up to first order leads to%
\begin{equation}
\begin{aligned}
\partial_t\,\delta\vektor{\xi}(y,t)=\: & \Big(\Jacobian+\partial_y^2\,\Tensor{D}-K\,\Cmat\Big)\,\delta\vektor{\xi}(y,t)\\
 & +K\,\Cmat\,\delta\vektor{\xi}(y,t-\tau)\:.
\end{aligned}
\label{eq:35}
\end{equation}

\begin{figure}[htb]
\centering
\includegraphics[scale=1]{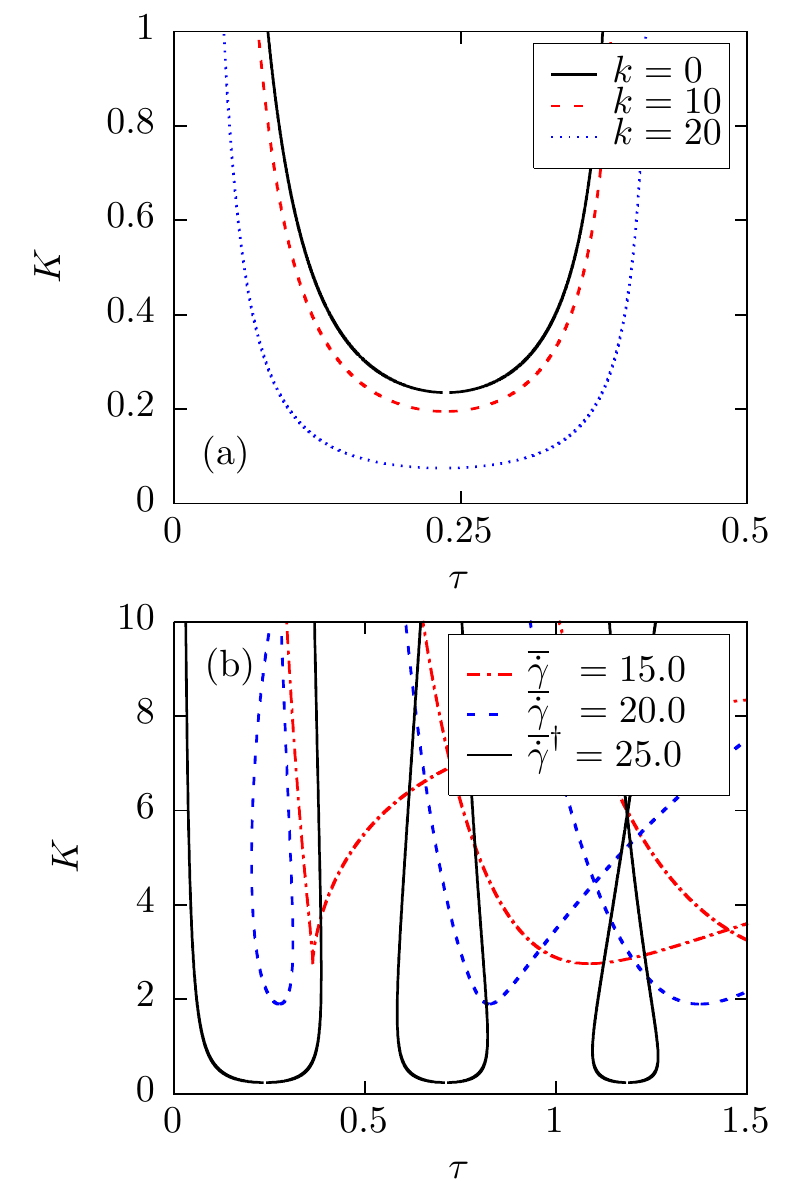}
\caption{(a) Stability borders for different wave numbers $k$ as functions of the delay time $\tau$ (at fixed $\overline{\dot{\gamma}}=25.0$ and $\tau_n=0.180$) and (b) neutral stability curves for $k=0$ at fixed $\tau_n=0.180$ for various shear rates. The dagger marks a shear rate close to the Hopf bifurcation.}
\label{fig:9}
\end{figure}

The eigenvalue problem then takes the form (with $\delta\vektor{\xi}(y,t)=\vektor{C}\,e^{\nu t+iky}$)%
\begin{equation}
\begin{vmatrix} \Jel_{11}-\nu & \Jel_{12}\\ \Jel_{21} & \Jel_{22}-k^2\,\mathcal{D}-K\,(1-e^{-\nu\tau})-\nu\end{vmatrix}=0\:,
\label{eq:36}
\end{equation}

where $\Jel_{ij}$ ($i,j=1,2$) are the components of the Jacobian matrix given in Eq.~(\ref{eq:26}). In principle, we have to distinguish here again between two cases, namely $k=0$ and $k>0$. In analogy to the uncontrolled case it can be shown that the most unstable mode occurs at $k=0$ (see also Fig. \ref{fig:9} (a)). Thus we consider here only the latter case, yielding%
\begin{equation}
\begin{aligned}
\nu^2 & +\nu\,\Big\lbrace-K\,\left(e^{-\nu\tau}-1\right)-A\Big\rbrace\\
 & +\Big\lbrace \Jel_{11}\,K\,\left(e^{-\nu\tau}-1\right)+B\Big\rbrace=0\:,
\end{aligned}
\label{eq:37}
\end{equation}

with $A=\Jel_{11}+\Jel_{22}$ and $B=\Jel_{11}\,\Jel_{22}-\Jel_{12}\,\Jel_{21}$.%

To simplify the analysis we consider a state close to the Hopf bifurcation, where the real $\mu$ part of the eigenvalue $\nu$ is approximately zero and thus, $\nu=i\,\omega$ is purely imaginary. Substituting this assumption into Eq.~(\ref{eq:37}) leads to%
\begin{equation}
\begin{aligned}
-\omega^2 & +i\omega\,\Big[-K\,\left(e^{-i\omega\tau}-1\right)-A\Big]\\
 & +\Big[\Jel_{11}\,K\,\left(e^{-i\omega\tau}-1\right)+B\Big]=0\:.
\end{aligned}
\label{eq:38}
\end{equation}

Separating Eq.~(\ref{eq:38}) into real and imaginary part yields
\begin{eqnarray}
\omega^2-K\Big[\Jel_{11}\cos\left(\omega\tau\right)-\omega\sin\left(\omega\tau\right)-\Jel_{11}\Big]-B & = & 0\:, \label{eq:39}\\
\omega A-K\Big[-\omega\cos\left(\omega\tau\right)-\Jel_{11}\sin\left(\omega\tau\right)+\omega\Big] & = & 0\:. \label{eq:40}
\end{eqnarray}

Solving Eqs.~(\ref{eq:39}) and (\ref{eq:40}) yields a pair of solutions $(\tau,K)$, which depend on the imaginary part $\omega$ of the eigenvalue $\nu$. Specifically, the delay time $\tau=\tau(\omega)$ is given by%
\begin{widetext}
\begin{equation}
\tau(\omega)=\dfrac{1}{\omega}\,\left[\arctan\left(\dfrac{2\omega\,\left(A\,\Jel_{11}-B+\omega^2\right)\,\left(B\,\Jel_{11}+\left(A-\Jel_{11}\right)\,\omega^2\right))}{(B\,\Jel_{11})^2-C_1\,\omega^2+C_2\,\omega^4-\omega^6}\right)+2j\pi\right]\,,\:\:j\in\mathbb{Z}
\label{eq:41}
\end{equation}
\end{widetext}

with $C_1=B^2+\left(A\,\Jel_{11}\right)^2+2B\,\Jel_{11}\,\left(\Jel_{11}-2A\right)$ and $C_2=A^2-2B-4A\,\Jel_{11}+\Jel_{11}^2$. These two constants depend on the system parameters $\alpha$, $\beta$, $\tau_n$, $\eta$ and $\overline{\dot{\gamma}}$. Further, the argument of the arctangent in Eq.~(\ref{eq:41}) is by definition limited to the interval $\left[-\pi/2,+\pi/2\right]$. The control strength $K=K(\omega)$ follows as%
\begin{equation}
K(\omega)=\dfrac{\left(A\,\omega\right)^2+\left(\omega^2-B\right)^2}{2\,\left(B\,\Jel_{11}+\omega^2\,\left(A-\Jel_{11}\right)\right)}\:.
\label{eq:42}
\end{equation}

These results are similar to the ones derived in Ref. \cite{Kyrychko2009} for the Gray-Scott model. In Fig. \ref{fig:9} (b) we plot the stability borders corresponding to the expressions for $\tau$ and $K$ in Eqs.~(\ref{eq:41}) and (\ref{eq:42}) for several shear rates at a fixed micellar relaxation time $\tau_n=0.180$. With increasing $\overline{\dot{\gamma}}$ we observe the minima of the stability curves to be shifted towards smaller delay times. Further, the width of the stability domains is increasing. Interestingly, the results in Fig. \ref{fig:9} (b) are similar to those obtained for a diagonal control scheme (see Refs. \cite{Strehober2013,Strehober2013a,Zeitz2015}), where a (local) Pyragas term is applied to both, the viscoelastic stress and the micellar length.%

\begin{figure}[htb]
\centering
\includegraphics[scale=1]{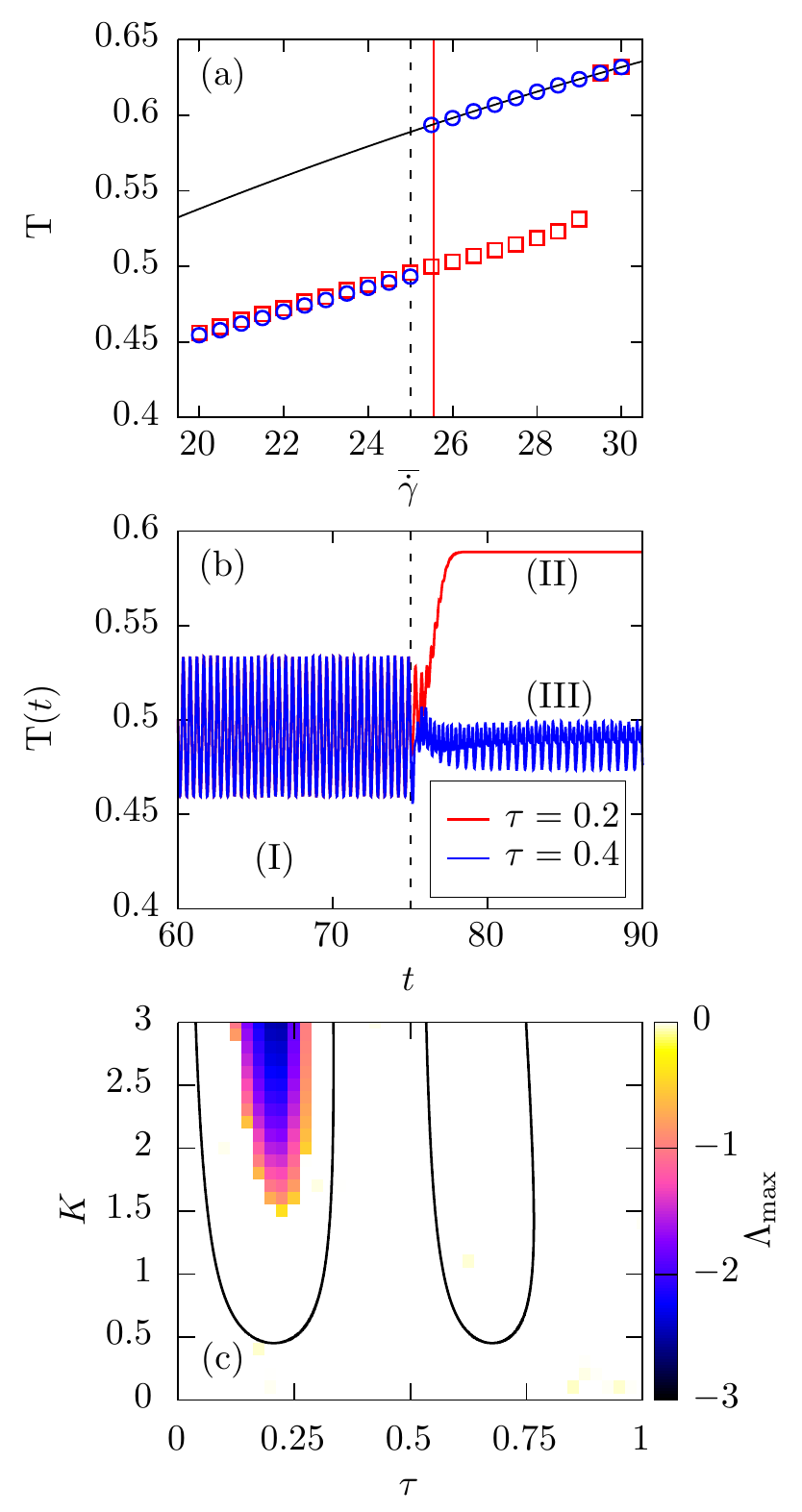}
\caption{(a) Time-averaged values of the total stress $\Ttot(\overline{\dot{\gamma}})$ as determined from numerics (squares and circles) compared to the analytical steady state solution (solid line, Eq.~(\ref{eq:14}), upper branch: stationary solution, lower branch: oscillatory solution). The vertical red line marks the Hopf bifurcation determined by linear stability analysis.  (b) Total stress $\Ttot(t)$ measured for two delay times $\tau$ at $K=3.0$ with the feedback control acting on the system for $t>75$ and (c) largest Lyapunov exponent $\Lambda_{\max}$ determined from numerics; the solid black lines represent the stability borders (fixed parameters: $\tau_=0.180$, $\overline{\dot{\gamma}}=25.0$, $\alpha=1.2$, $\beta=1.5$, $\eta=0.005$, $\mathcal{D}=0.0016$).}
\label{fig:10}
\end{figure}

As indicated at the beginning of Sec. \ref{sec:IV_A}, we apply the feedback control to states close to the (analytical) Hopf bifurcation (red line in Fig. \ref{fig:10} (a)), where the system reveals a hysteretic behaviour as shown in Fig. \ref{fig:10} (a). Specifically, for fixed $\overline{\dot{\gamma}}$, we find the numerical solution for the total stress to have two solution branches, one corresponding to the (constant) steady state solution $\Ttot_s(\overline{\dot{\gamma}})$ determined in Eq.~(\ref{eq:14}), and the other one to a solution where the total stress fluctuates in time (the solution plotted in Fig. \ref{fig:10} (a) corresponds to the time average).

In order to see the impact of the (non-diagonal) control scheme on the spatio-temporal dynamics, we record the total stress $\Ttot(t)$ (exemplary shown in Fig. \ref{fig:10} (b) for a fixed control strength and two delay times). In the uncontrolled case ($t<75$) the system is in an oscillatory state (I). When the control term is switched on ($t\ge75$) we find two scenarios: The initial state (I) can be either fully stabilized (II), i.e. the total stress reaches a constant value corresponding to the steady state solution $\Ttot_s(\overline{\dot{\gamma}})$ in the long time limit, or the control scheme is not working (III) and the oscillating character of the initial state is preserved, but the amplitude of the oscillations is damped. 

We stress that, in case of reaching the steady state after applying the feedback control, the solution is not chosen arbitrarily, but is already one of the possible solutions (namely the stationary branch) given by the uncontrolled system (due to the hysteretic behaviour shown in Fig. \ref{fig:10} (a)). This selection of an existing solution underlines the non-invasive character of the Pyragas scheme.

\begin{figure}[htb]
\centering
\includegraphics[scale=1]{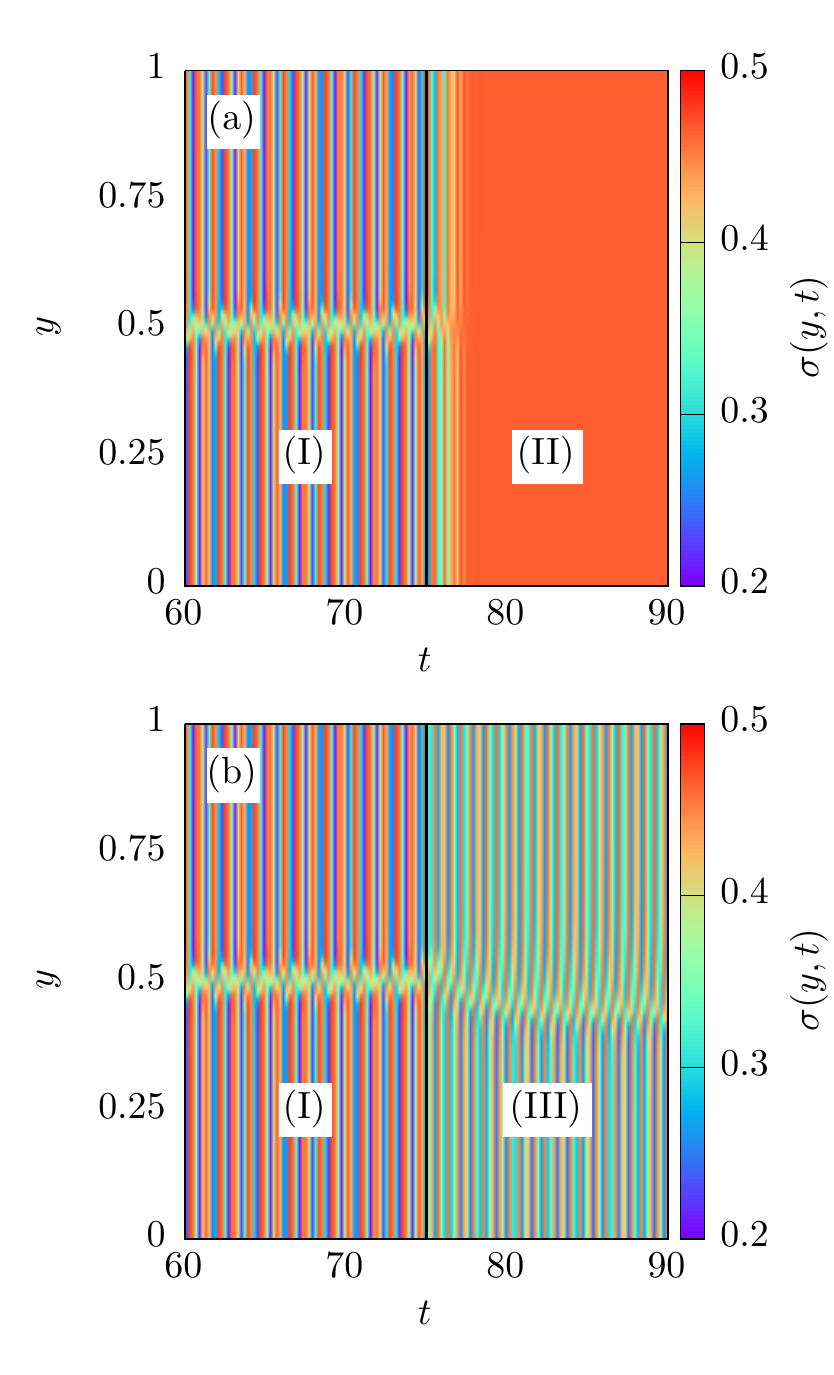}
\caption{Space-time plots of the viscoelastic stress $\sigma(y,t)$ for (a) $\tau=0.2$ and (b) $\tau=0.4$ at $K=3.0$. The feedback control acts on the system for $t>75$ (fixed parameters: $\tau_=0.180$, $\overline{\dot{\gamma}}=25.0$, $\alpha=1.2$, $\beta=1.5$, $\eta=0.005$, $\mathcal{D}=0.0016$).}
\label{fig:11}
\end{figure}

From the time series $\Ttot(t)$ we calculate the largest Lyapunov exponent $\Lambda_{\max}$ by using the Wolf-algorithm \cite{Wolf1985}. The results for $\Lambda_{\max}$ are presented in Fig. \ref{fig:10} (c). In case of successful control we find $\Lambda_{\max}<0$ (corresponding to region (II) in Fig. \ref{fig:10} (b)). If the control scheme is not successful, the Lyapunov exponent is positive ($\Lambda_{\max}>0$). To visualize the impact of the feedback control on the dynamics of the system we present in Fig. \ref{fig:11} the spatio-temporal evolution of $\sigma(y,t)$, which correspond to the time series of the total stress presented in Fig. \ref{fig:10} (b). The situation shown in Fig. \ref{fig:11} (a) refers to a successful application of the non-diagonal TDFC scheme, i.e. a spatio-temporal inhomogeneous state is stabilized to a homogeneous steady state, while in Fig. \ref{fig:11} (b) the control scheme does not work in the sense that the inhomogeneous character of the initial state is preserved.%

\subsection{Stress controlled protocol}\label{sec:IV_B}

\begin{figure}
\centering
\includegraphics[scale=1]{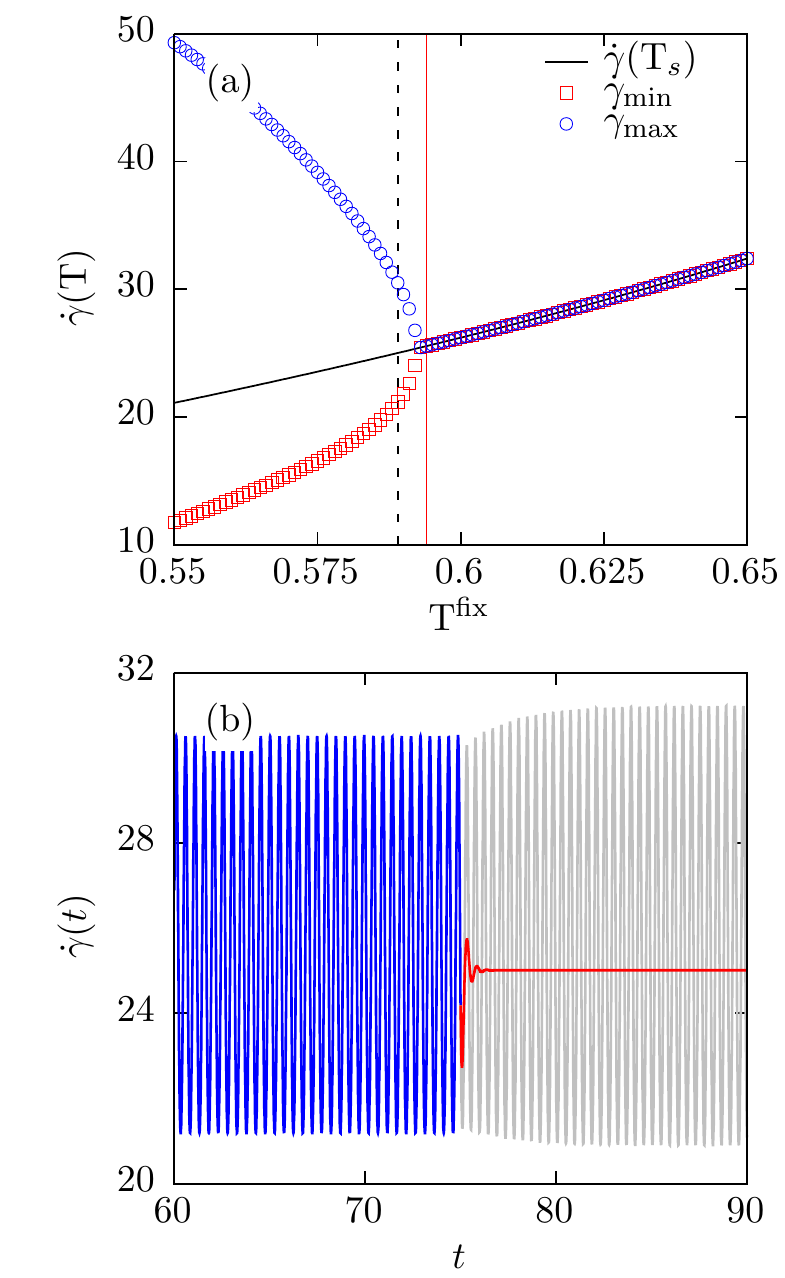}
\caption{(a) Shear rate $\dot{\gamma}$ as function of the applied stress $\Ttot^{\mathrm{fix}}$. The symbols represent the minimum and maximum of the oscillations of $\dot{\gamma}$. (b) Shear rate $\dot{\gamma}(t)$ measured at $\Ttot^{\mathrm{fix}}=0.589$ for the uncontrolled system (blue line, $t<75$) and TDFC applied to the system for $t>75$ with $(\tau,K^\ast)=(0.2,3.0)$ (red line) and $(\tau,K^\ast)=(0.4,3.0)$ (grey line) (fixed parameters: $\tau_n=0.180$, $\alpha=1.2$, $\beta=1.5$, $\eta=0.005$).}
\label{fig:12}
\end{figure}

Similarly to the shear rate controlled protocol, we can also construct a Pyragas-like feedback term within the stress controlled protocol. The corresponding Pyragas-like term to be added in the equation of motion has now the form
\begin{equation}
\begin{aligned}
\dot{\gamma}(t)-\dot{\gamma}(t-\tau) & =\dfrac{1}{\eta}\,\Big[\Ttot^{\mathrm{fix}}-\sigma(t)-\Ttot^{\mathrm{fix}}+\sigma(t-\tau)\Big]\\
 & =-\dfrac{1}{\eta}\,\Big[\sigma(t)-\sigma(t-\tau)\Big]\:,
\end{aligned}
\label{eq:43}
\end{equation}

where we used the definition of the shear rate in the stress controlled protocol given in Eq.~(\ref{eq:8}). Henceforth, we focus on the non-diffusive limit with $\mathcal{D}=0$, following the arguments presented in Sec. \ref{sec:II_C}. We again restrict ourselves to states close to the Hopf bifurcation, where the system reveals temporal oscillations. Within this regime we find the shear rate oscillating around its steady state values (see Fig. \ref{fig:12} (a)). We will show that the feedback-controlled system (analogous to the situation considered in the previous section) selects this steady state solution for certain sets of control parameters $K$ and $\tau$.

The form of Eq.~(\ref{eq:43}) already suggests that the feedback control should again be applied as a non-diagonal scheme, i.e. the TDFC is only acting on the viscoelastic stress. Using the vector notation we introduced in Eq.~(\ref{eq:29}) for the stress controlled protocol, we can formulate a (generalized) feedback equation such that
\begin{equation}
\partial_t\,\vektor{\xi}(t)=\vektor{Q}\left(\vektor{\xi},\Ttot^{\mathrm{fix}}\right)-K^\ast\,\Cmat\,\Big[\vektor{\xi}(t)-\vektor{\xi}(t-\tau)\Big]\:,
\label{eq:44}
\end{equation}

where $\Cmat$ is the matrix defined in Eq.~(\ref{eq:34}) ensuring that the TDFC is only acting on the viscoelastic stress. Taking Eq.~(\ref{eq:43}) into account yields the specific feedback equation
\begin{equation}
\partial_t\,\vektor{\xi}(t)=\vektor{Q}\left(\vektor{\xi},\Ttot^{\mathrm{fix}}\right)+\dfrac{K}{\eta}\,\Cmat\,\Big[\vektor{\xi}(t)-\vektor{\xi}(t-\tau)\Big]\:.
\label{eq:45}
\end{equation}

Comparing Eqs.~(\ref{eq:44}) and (\ref{eq:45}) we identify $K^\ast=-K/\eta$ as generalized feedback strength. Due to reasons of comparability with the previous section, we focus in the following on the generalized delay differential equation~(\ref{eq:44}). To determine the stability domains of Eq.~(\ref{eq:44}) we follow the same procedure as outlined in Sec. \ref{sec:IV_A} for the shear rate controlled protocol. Since we consider the non-diffusive limit, the perturbations $\delta\vektor{\xi}(t)=\vektor{C}\,e^{\nu t}$ with $\vektor{C}=(C_1,C_2)^T$ are only time-dependent. Further, the Jacobian matrix now depends on the fixed total stress $\Ttot^{\mathrm{fix}}$ (see also Appendix \ref{Appendix_B}).

Interestingly, we find the shape of the neutral stability curves to be unchanged in comparison to the ones of the shear rate controlled protocol depicted in Fig. \ref{fig:9} (b). The reason for this behaviour can be found in the coupling between the shear rate $\dot{\gamma}$ and the total stress $\Ttot$ given in Eq.~(\ref{eq:14}), i.e. the shear rates chosen in Fig. \ref{fig:9} (b) can be converted to the corresponding stresses. 

More interesting is the impact of the TDFC scheme on the dynamics of stress controlled protocol, which is depicted in Fig. \ref{fig:12} (b). Measuring the shear rate as function of time we find the uncontrolled system to reveal oscillations, which are damped out if the control is activated with $(\tau,K^\ast)=(0.2,3.0)$ for times $t>75$ (red lines in Fig. \ref{fig:12} (b)). The resulting constant value is the same as the one determined from the steady state solution (see Fig. \ref{fig:11} (a)). This again emphasizes the non-invasive character of the Pyragas scheme, i.e. the system takes an already existing solution when applying TDFC. On the other hand for the parameter set $(\tau,K^\ast)=(0.4,3.0)$ the oscillations are preserved, but the amplitude is changed (grey lines in Fig. \ref{fig:12} (b)).

\section{Conclusions}

In this publication we have applied time-delayed feedback control (specifically, the Pyragas scheme) to manipulate the complex dynamics of a solution of wormlike micelles based on a non-local rheological model proposed by Fielding and Olmsted \cite{Fielding2004}. As a background we have performed a linear stability analysis (at $k=0$) for uncontrolled systems at constant (average) shear rate, as well as systems at constant total stress. Further, we solved the dynamical system numerically to get an insight into the full spatio-temporal behaviour for a broad range of parameters. In doing this, we considered both an ``up'' ramp, i.e. increasing the shear from low values, and a ``down'' ramp, that is, decreasing the shear rate from high values. Comparing the corresponding numerical results with the analytical predictions from linear stability analysis (at $k=0$), we find a good agreement of the data for the ``down'' ramp. For the ``up'' ramp, the system displays hysteric behaviour not captured by the linear stability analysis. This hysteresis effect has already been observed in the Johnson-Segalman model \citep{Adams2008} within the bistable region related to shear banding. In the model investigated here, the hysteresis occurs near the Hopf bifurcation located at the high shear rate branch of the constitutive curve.%

We have then investigated the changes of the dynamics upon application of a global and local Pyragas control scheme \cite{Pyragas1992}, where the control variable is associated with a measurable quantity of the corresponding rheological protocol. In case of the shear rate controlled protocol, we have shown that a local control scheme for the stress can stabilize the spatio-temporal oscillatory dynamics close to the Hopf bifurcation, where the system exhibits hysteresis. Specifically, by applying feedback control of suitable strength and delay time, the system selects the steady state solution branch. In contrast, a global scheme acting only on the spatially averaged stress does not work. However, a global scheme is indeed successful within the stress controlled protocol, where the dynamics is only time dependent. Again we only focused on states close to the analytically determined Hopf bifurcation, where the system reveals temporal oscillations. Here, the steady state solution is again part of the full solution space of the system. In analogy to the shear rate controlled protocol, we have shown that the feedback control can select the steady state solution branch. This again underlines the non-invasive character of the Pyragas scheme.%

We are aware that the control scheme presented here is not generally applicable to all shear-induced states. The reason is that, outside the hysteretic regions considered here, the relation between $\Ttot$ and $\overline{\dot{\gamma}}$ is unique, that is, one cannot control one of these variables without affecting the other one. Therefore, it is reasonable to think of alternative ways of implementing feedback control to sheared systems. On possibility might be a \emph{externa}l (mechanical) control, where the stress $\Ttot$ or the shear rate $\dot{\gamma}$ is serving as a measurable quantity, which is fed back to the system using a transducer manipulating the (externally) imposed variable. To implement such kind of feedback the (full) Navier-Stokes equation is required to get access to the velocity profile and therewith define a coupling to a transducer. Another possible scenario would be the implementation of an \emph{internal} control, acting on the structural variables: Hereby, the externally imposed rheological variable ($\overline{\dot{\gamma}}$ or $\Ttot^{\mathrm{fix}}$) is kept constant and the structural properties of the system are changed by optical, electric or magnetic fields, that is, the conformation or the stiffness of aggregates and molecules will be tuned through interactions with an external field. In this case, a further dynamical equation for the external field has to be defined. Clearly, it would be very interesting to test the implications of such feedback control (or related schemes) experimentally. Especially the internal control might be relevant in view of the increasing importance of stimuli responsive materials \cite{Tsitsilianis2010,Chu2013,Brassinne2015}.

We also mention recent progress in accessing the spatio-temporal structure of micellar solutions under shear using appropriate scattering techniques (Rheo-SANS) \cite{Helgeson2009,Lopez-Barron2014}. Given these developments, an application of TDFC to avoid or stabilize certain shear-induced behaviours seems very promising. Further, it seems to be interesting to systematically vary the stress-diffusion constant $\mathcal{D}$ and the power law exponents $\alpha$ and $\beta$ to determine the impact on the dynamics of the system and to compare them to experimental data.

\begin{acknowledgements}
We gratefully acknowledge financial support from the Deutsche Forschungsgemeinschaft (DFG) through project B2 within the Collaborative Research Center 910. The authors thank M. Gradzielski for fruitful discussions and the insight from the experimental point of view.
\end{acknowledgements}

\appendix

\section{Jacobian matrix for the shear rate controlled protocol} \label{Appendix_A}

We start from the compact notation for the dynamical system given in Eq.~(\ref{eq:19}) in Sec. \ref{sec:III_A_1}
\begin{equation}
\partial_t\,\vektor{\xi}(y,t)=\vektor{Q}\left(\vektor{\xi},\dot{\gamma}\right)+\partial_y^2\,\left(\Tensor{\mathcal{D}}\,\vektor{\xi}(y,t)\right)\:,
\label{eq:A1}
\end{equation}

where $\Tensor{\mathcal{D}}$ is the diffusion matrix and the vector $\vektor{Q}\left(\vektor{\xi},\dot{\gamma}\right)$ follows from Eq.~(\ref{eq:9}) and (\ref{eq:11})
\begin{equation}
\begin{aligned}
\vektor{Q}\left(\vektor{\xi},\dot{\gamma}\right) & =\begin{pmatrix}-\frac{n(y,t)}{\tau_n}+\frac{1}{\tau_n}\,\left(\frac{n_0}{1+\left(\tau_n\,\dot{\gamma}(y,t)\right)^\beta}\right)\\ -\frac{\sigma(y,t)}{\tau(n)}+\frac{\dot{\gamma}(y,t)}{1+\left(\tau(n)\,\dot{\gamma}(y,t)\right)^2}\end{pmatrix}\\
 & =\begin{pmatrix}Q_1\left(n,\sigma,\dot{\gamma}\right)\\ Q_2\left(n,\sigma,\dot{\gamma}\right)\end{pmatrix}\:.
\end{aligned}
\label{eq:A2}
\end{equation}

The Jacobian matrix $\Jacobian$ for the shear rate controlled protocol is defined by
\begin{equation}
\Jacobian=\Mmat-\dfrac{1}{\eta}\,\vektor{q}\cdot\vektor{p}\:,
\label{eq:A3}
\end{equation}

with $\Mmat=\left.\partial_{\vektor{\xi}}\,\vektor{Q}\right|_{\vektor{\xi}_s,\overline{\dot{\gamma}}}$ and $\vektor{q}=\left.\partial_{\dot{\gamma}}\,\vektor{Q}\right|_{\vektor{\xi}_s,\overline{\dot{\gamma}}}$. We first consider the matrix $\Mmat$. The latter can be written in terms of the components of the vector $\vektor{Q}$
\begin{equation}
\Mmat=\left.\partial_{\vektor{\xi}}\,\vektor{Q}\right|_{\vektor{\xi}_s,\overline{\dot{\gamma}}}=\left.\begin{pmatrix}\partial_n\,Q_1 & \partial_\sigma\,Q_1\\ \partial_n\,Q_2 & \partial_\sigma\,Q_2\end{pmatrix}\right|_{\vektor{\xi}_s,\overline{\dot{\gamma}}}\:.
\label{eq:A4}
\end{equation}

Inserting the expressions for $Q_1\left(n,\sigma,\dot{\gamma}\right)$ and $Q_2\left(n,\sigma,\dot{\gamma}\right)$ given in Eq.~(\ref{eq:A2}) and evaluating the derivatives at the steady state values one obtains
\begin{eqnarray}
\left.\partial_n\,Q_1\left(n,\sigma,\dot{\gamma}\right)\right|_{\vektor{\xi}_s,\overline{\dot{\gamma}}} & = & -\dfrac{1}{\tau_n} \label{eq:A5}\\
\left.\partial_\sigma\,Q_1\left(n,\sigma,\dot{\gamma}\right)\right|_{\vektor{\xi}_s,\overline{\dot{\gamma}}} & = & 0 \label{eq:A6}\\
\left.\partial_n\,Q_2\left(n,\sigma,\dot{\gamma}\right)\right|_{\vektor{\xi}_s,\overline{\dot{\gamma}}} & = & \dfrac{\alpha\,\sigma_s}{n_s^{\alpha+1}}-\dfrac{2\alpha\,\overline{\dot{\gamma}}^3\,n_s^{2\alpha-1}}{\left[1+\left(\overline{\dot{\gamma}}\,n_s^\alpha\right)^2\right]^2}  \label{eq:A7}\\
\left.\partial_\sigma\,Q_2\left(n,\sigma,\dot{\gamma}\right)\right|_{\vektor{\xi}_s,\overline{\dot{\gamma}}} & = & -\dfrac{1}{n_s^\alpha}\:. \label{eq:A8}
\end{eqnarray}

Second, we consider the vector $\vektor{q}$,
\begin{equation}
\vektor{q}=\partial_{\dot{\gamma}}\,\left.\vektor{Q}\right|_{\vektor{\xi}_s,\overline{\dot{\gamma}}}=\left.\begin{pmatrix}\partial_{\dot{\gamma}}\,Q_1\\ \partial_{\dot{\gamma}}\,Q_2\end{pmatrix}\right|_{\vektor{\xi}_s,\overline{\dot{\gamma}}}\:.
\label{eq:A9}
\end{equation}

Using Eq.~(\ref{eq:A2}) we obtain
\begin{eqnarray}
\left.\partial_{\dot{\gamma}}\,Q_1\left(n,\sigma,\dot{\gamma}\right)\right|_{\vektor{\xi}_s,\overline{\dot{\gamma}}} & = & -\dfrac{\beta\,\left(\tau_n\,\overline{\dot{\gamma}}\right)^\beta}{\tau_n\,\overline{\dot{\gamma}}\,\left[1+\left(\tau_n\,\overline{\dot{\gamma}}\right)^\beta\right]^2}\:, \label{eq:A10}\\
\left.\partial_{\dot{\gamma}}\,Q_2\left(n,\sigma,\dot{\gamma}\right)\right|_{\vektor{\xi}_s,\overline{\dot{\gamma}}} & = & \dfrac{1-\left(\overline{\dot{\gamma}}\,n_s^\alpha\right)^2}{\left[1+\left(\overline{\dot{\gamma}}\,n_s^\alpha\right)^2\right]^2}\:. \label{eq:A11}
\end{eqnarray}

Inserting the elements of $\Mmat$ and $\vektor{q}$ into Eq.~(\ref{eq:A3}) for the matrix $\Jacobian$, one finds the following elements $\Jel_{ij}$ ($i,j=1,2$)
\begin{eqnarray}
\Jel_{11} & = & -\dfrac{1}{\tau_n} \label{eq:A12}\\
\Jel_{12} & = & \dfrac{\beta\,\left(\tau_n\,\overline{\dot{\gamma}}\right)^\beta}{\eta\,\tau_n\,\overline{\dot{\gamma}}\,\left[1+\left(\tau_n\,\overline{\dot{\gamma}}\right)^\beta\right]^2} \label{eq:A13}\\
\Jel_{21} & = & \dfrac{\alpha\,\sigma_s}{n_s^{\alpha+1}}-\dfrac{2\alpha\,\overline{\dot{\gamma}}^3\,n_s^{2\alpha-1}}{\left[1+\left(\overline{\dot{\gamma}}\,n_s^\alpha\right)^2\right]^2} \label{eq:A14}\\
\Jel_{22} & = & -\dfrac{1}{n_s^\alpha}-\dfrac{1-\left(\overline{\dot{\gamma}}\,n_s^\alpha\right)^2}{\eta\,\left[1+\left(\overline{\dot{\gamma}}\,n_s^\alpha\right)^2\right]^2}\:. \label{eq:A15}
\end{eqnarray}

For convenience we set $n_0=\tau_0=1.0$ for all calculations.

\section{Jacobian matrix for the stress controlled protocol} \label{Appendix_B}

In the stress controlled protocol the shear rate is replaced by $\dot{\gamma}(y,t)=1/\eta\,\left[\Ttot^{\mathrm{fix}}-\sigma(y,t)\right]$, with $\Ttot^{\mathrm{fix}}$ being a constant. In vector notation the dynamical system can be written
\begin{equation}
\partial_t\,\vektor{\xi}(y,t)=\vektor{Q}\left(\vektor{\xi},\Ttot^{\mathrm{fix}}\right)\:,
\label{eq:B1}
\end{equation}

where the vector $\vektor{Q}\left(\vektor{\xi},\Ttot^{\mathrm{fix}}\right)$ is given by
\begin{equation}
\begin{aligned}
\vektor{Q}\left(\vektor{\xi},\Ttot^{\mathrm{fix}}\right)= & \begin{pmatrix}-\frac{n(y,t)}{\tau_n}+\frac{1}{\tau_n}\,\left(\frac{n_0}{1+\left(\tau_n/\eta\,\left[\Ttot^{\mathrm{fix}}-\sigma(y,t)\right]\right)^\beta}\right)\\ -\frac{\sigma(y,t)}{\tau\left(n\right)}+\frac{1/\eta\,\left[\Ttot^{\mathrm{fix}}-\sigma(y,t)\right]}{1+\left[\tau\left(n\right)/\eta\,\left[\Ttot^{\mathrm{fix}}-\sigma(y,t)\right]\right]^2}\end{pmatrix}\\
 & =\begin{pmatrix}Q_1\left(n,\sigma,\Ttot^{\mathrm{fix}}\right)\\ Q_2\left(n,\sigma,\Ttot^{\mathrm{fix}}\right)\end{pmatrix}\:.
\end{aligned}
\label{eq:B2}
\end{equation}

The Jacobian matrix $\Jacobian$ evaluated at the steady state values can be written as
\begin{equation}
\Jacobian=\left.\partial_{\vektor{\xi}}\,\vektor{Q}\right|_{\vektor{\xi}_s,\Ttot^{\mathrm{fix}}}=\left.\begin{pmatrix}\partial_n\,Q_1 & \partial_\sigma\,Q_1\\ \partial_n\,Q_2 & \partial_\sigma\,Q_2\end{pmatrix}\right|_{\vektor{\xi}_s,\Ttot^{\mathrm{fix}}}\:.
\label{eq:B3}
\end{equation}

Explicitly, the components $\Jel_{ij}$ ($i,j=1,2$) of the Jacobian matrix are given by
\begin{eqnarray}
\Jel_{11} & = & -\dfrac{1}{\tau_n} \label{eq:B4}\\
\Jel_{12} & = & \dfrac{\beta\,\left[\tau_n/\eta\,\left(\Ttot^{\mathrm{fix}}-\sigma_s\right)\right]^{\beta-1}}{\eta\,\left[1+\left(\tau_n/\eta\,\left(\Ttot^{\mathrm{fix}}-\sigma_s\right)\right)^\beta\right]^2} \label{eq:B5}\\
\Jel_{21} & = & \dfrac{\alpha}{n_s^{\alpha+1}}\,\left[\sigma_s+\dfrac{2\eta\,\left(n_s^\alpha\,\left(\Ttot^{\mathrm{fix}}-\sigma_s\right)\right)^3}{\left[\eta^2+\left(n_s^\alpha\,\left(\Ttot^{\mathrm{fix}}-\sigma_s\right)\right)^2\right]^2}\right] \label{eq:B6}\\
\Jel_{22} & = & -\dfrac{1}{n_s^\alpha}+\dfrac{\eta\,\left[\left(n_s^\alpha\,\left(\Ttot^{\mathrm{fix}}-\sigma_s\right)\right)^2-\eta^2\right]}{\left[\left(n_s^\alpha\,\left(\Ttot^{\mathrm{fix}}-\sigma_s\right)\right)^2+\eta^2\right]^2}\:. \label{eq:B7}
\end{eqnarray}

For convenience we set $n_0=\tau_0=1.0$ for all calculations.

\section{Linear stability of the global feedback scheme in the shear rate controlled protocol} \label{Appendix_C}

We start our derivation of the linear stability analysis of the global feedback control scheme in the shear rate controlled protocol with the dynamical equations in vector notation (analogous to Eq.~(\ref{eq:33})), that is
\begin{equation}
\begin{aligned}
\partial_t\,\vektor{\xi}(y,t)= & \:\vektor{Q}\left(\vektor{\xi},\dot{\gamma}\right)+\partial_y^2\,\left(\Tensor{\mathcal{D}}\,\vektor{\xi}(y,t)\right)\\
 & \:-K\,\Cmat\,\Big[\overline{\vektor{\xi}}(t)-\overline{\vektor{\xi}}(t-\tau)\Big]\:,
\end{aligned}
\label{eq:C1}
\end{equation}

where $\Tensor{\mathcal{D}}$ is the diffusion matrix defined in Eq.~(\ref{eq:20}) and $\Cmat$ is a non-diagonal $2\times2$ matrix defined in Eq.~(\ref{eq:34}) ensuring that the feedback control is only acting on the viscoelastic stress $\sigma$. Further, $\overline{\vektor{\xi}}$ represents the spatially averaged version of the dynamical vector $\vektor{\xi}(y,t)=\left[n(y,t),\sigma(y,t)\right]^T$. Decomposing the dynamical variable $\vektor{\xi}(y,t)$ as well as the spatially averaged variable $\overline{\vektor{\xi}}(t)$ into a steady state plus small perturbations, substituting this back to Eq.~(\ref{eq:C1}) and linearising in the perturbations up to first order gives rise to the following equation
\begin{equation}
\begin{aligned}
\partial_t\,\delta\vektor{\xi}(y,t)= & \Big(\Jacobian+\partial_y^2\,\Tensor{\mathcal{D}}\Big)\,\delta\vektor{\xi}(y,t)\\
 & -K\,\Cmat\,\Big[\delta\overline{\vektor{\xi}}(t)-\delta\overline{\vektor{\xi}}(t-\tau)\Big]\:,
\end{aligned}
\label{eq:C2}
\end{equation}

where $\Jacobian$ is the Jacobian matrix derived in Appendix \ref{Appendix_A}. For the perturbations $\delta\vektor{\xi}(y,t)$ we choose the following ansatz
\begin{equation}
\delta\vektor{\xi}(y,t)=\vektor{C}\,e^{\nu t+iky}\:,
\label{eq:C3}
\end{equation}

with $\vektor{C}=(C_1,C_2)^T$ being an arbitrary constant vector, $\nu$ the complex eigenvalues and $k$ the wave number. Averaging the perturbations in Eq.~(\ref{eq:C2}) over space yields
\begin{equation}
\begin{aligned}
\delta\overline{\vektor{\xi}}(t) & =\dfrac{1}{L}\,\int\limits_0^L\mathrm{d}y\,\delta\vektor{\xi}(y,t)=\dfrac{\vektor{C}\,e^{\nu t}}{L}\,\int\limits_0^L\mathrm{d}y\,e^{iky}\\
 & =\vektor{C}\,e^{\nu t}\,\delta_{k,0}\:.
\end{aligned}
\label{eq:C4}
\end{equation}

Hereby $\delta_{k,0}$ is the Kronecker-delta. Substituting Eqs.~(\ref{eq:C3}) and (\ref{eq:C4}) back in Eq.~(\ref{eq:C2}) leads to
\begin{equation}
\begin{aligned}
\vektor{C}\,\nu\,e^{\nu t+iky}= & \Big(\Jacobian+\partial_y^2\,\Tensor{\mathcal{D}}\Big)\,\vektor{C}\,e^{\nu t+iky}\\
 & -K\,\Cmat\,\vektor{C}\,e^{\nu t}\,\delta_{k,0}\,\Big[1-e^{-\nu\tau}\Big]\:.
\end{aligned}
\label{eq:C5}
\end{equation}

This equation can be transferred to
\begin{equation}
\Big(\Jacobian+\partial_y^2\,\Tensor{\mathcal{D}}-\nu\,\Imat\Big)\,e^{iky}-K\,\Cmat\,\,\delta_{k,0}\,\Big[1-e^{-\nu\tau}\Big]=0\:.
\label{eq:C6}
\end{equation}

We have to distinguish two cases: For $k=0$ Eq.~(\ref{eq:C6}) is reduced to the eigenvalue problem we described for the local TDFC in Sec. \ref{sec:IV_A}. The case of $k\ne0$ yields a reduction to the eigenvalue problem of the system without feedback control as discussed in Sec. \ref{sec:III_A_1}. This implies that only the long wave limit can be stabilized, i.e. general validity for all wave numbers $k$ is not guaranteed, which is a necessary condition for linear stability analysis. We also confirmed these results by numerical test calculations.
\bibliography{references}{}

\begin{thebibliography}{60}
\bibitem[{\citenamefont{{A more detailed numerical investigation (MATCONT)
  reveals that this critical point is a Bogdanov-Takens point.}}()}]{critpoint}
\bibinfo{author}{\bibnamefont{{A more detailed numerical investigation
  (MATCONT) reveals that this critical point is a Bogdanov-Takens point.}}}
\end{thebibliography}
\bibliographystyle{apsrev}

\end{document}